\pdfoutput=1
\documentclass[twocolumn,showpacs,preprintnumbers,superscriptaddress,prb,amsmath,amssymb]{revtex4-1}

\usepackage{graphicx}
\usepackage{dcolumn}
\usepackage{bm}
\usepackage{color}
\usepackage{ulem}
\usepackage{gensymb}
\usepackage{braket}
\usepackage{epstopdf}
\usepackage{amsmath}
\usepackage{xcolor}
\usepackage{mathtools}
\usepackage[percent]{overpic}
\usepackage{enumitem}

\begin{document}

\definecolor{myred}{RGB}{154,29,29}
\definecolor{mygreen}{RGB}{29,140,29}
\definecolor{myred2}{RGB}{191,0,0}
\definecolor{mygreen2}{RGB}{0,169,0}
\definecolor{myorange1}{RGB}{237,144,0}

\newcommand{\cre}[2]{{#1}_{#2}^{\dagger}}
\newcommand{\ann}[2]{{#1}_{#2}^{\vphantom{\dagger}}}
\newcommand{\veck}[1]{\boldsymbol{#1}}

\newcommand{\dif}{\mathrm{d}}
\newcommand{\pard}[1]{\frac{\partial}{\partial #1}}
\newcommand{\pcre}[1]{\cre{\phi}{#1}}
\newcommand{\pann}[1]{\ann{\phi}{#1}}

\newcommand{\coloronline}{}

\title{Effective magnetic interactions in spin-orbit coupled $d^4$ Mott insulators}

\author{Christopher Svoboda}
\author{Mohit Randeria}
\author{Nandini Trivedi}
\affiliation{Department of Physics, The Ohio State University, Columbus, Ohio 43210, USA}
\date{\today}

\begin{abstract}
Transition metal compounds with the $(t_{2g})^4$ electronic configuration are expected to be nonmagnetic atomic singlets both in the weakly interacting regime due to spin-orbit coupling, as well as in the Coulomb dominated regime with oppositely aligned $L=1$ and $S=1$ angular momenta. 
However, starting with the full multi-orbital electronic Hamiltonian, we show the low energy effective magnetic Hamiltonian contains isotropic superexchange spin interactions but anisotropic orbital interactions. 
By tuning the ratio of superexchange to spin-orbit coupling $J_\mathrm{SE}/\lambda$, we obtain a phase transition from nonmagnetic atomic singlets to novel magnetic phases depending on the strength of Hund's coupling, the crystal structure and the number of active orbitals.
Spin-orbit coupling plays a non-trivial role in generating a triplon condensate of weakly interacting excitations at antiferromagnetic ordering vector $\vec k=\vec \pi$, regardless of whether the local spin interactions are ferromagnetic or antiferromagnetic.
In the large $J_\mathrm{SE} / \lambda$ regime, the localized spin and orbital moments produce anisotropic orbital interactions that are frustrated or constrained even in the absence of geometric frustration.
Orbital frustration leads to frustration in the spin channel opening up the possibility of spin-orbital liquids with both spin and orbital entanglement.
\end{abstract}

\pacs{
75.30.Et,
71.70.Ej,
75.10.Jm 
}

\maketitle

\section{Introduction}
Between weakly correlated topological insulators and strongly correlated 3d transition metal oxides lie $5d$ compounds that combine both strong spin-orbit coupling and correlations on an equal footing. 
In contrast to the well studied $5d^5$ materials, the effect of strong spin-orbit coupling in $4d^4$ and $5d^4$ systems have been sparsely studied due to expectations that these will naturally lead to non-magnetic insulating behavior.
However there are several experimental counter examples to this notion.
The first example is Ca$_2$RuO$_4$ which displays a moment of 1.3$\mu_\mathrm{B}$.\cite{Nakatsuji1997,Braden1998}
Recently both double perovskite irridates\cite{CaoKaul2013,MarcoHaskel2015,Dasgupta2015} and honeycomb ruthenates\cite{WangCao2014} in the $d^4$ configuration have been found to show magnetism.
It has been argued that partial quenching of the orbital angular momentum from the presence of lattice distortions is the root cause.
Very recent developments\cite{Kennedy2015,PhelanCava2016,Buchner2016,AritaKunes2016} on Ba$_2$YIrO$_6$ have piqued interest on the origin of magnetism in this $5d^4$ system because the compound is negligibly distorted and still shows a Curie response.

\begin{figure}
\vspace{0.1cm}
\begin{center}
\begin{overpic}[width=2.6in]{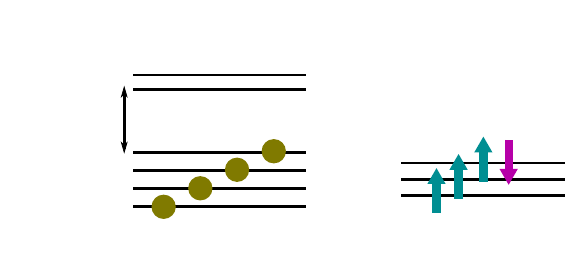}
 \put(-15,42.5){\textbf{(a)}}
 \put(25,42){$jj$ coupling}
 \put(69,42){$LS$ coupling}
 \put(2,15){$j=3/2$}
 \put(2,32){$j=1/2$}
 \put(71,28){$S=L=1$}
 \put(31.5,3){$J=0$}
 \put(77,3){$J=0$}
\end{overpic}
\end{center}
\begin{center}
\begin{overpic}[width=2.6in]{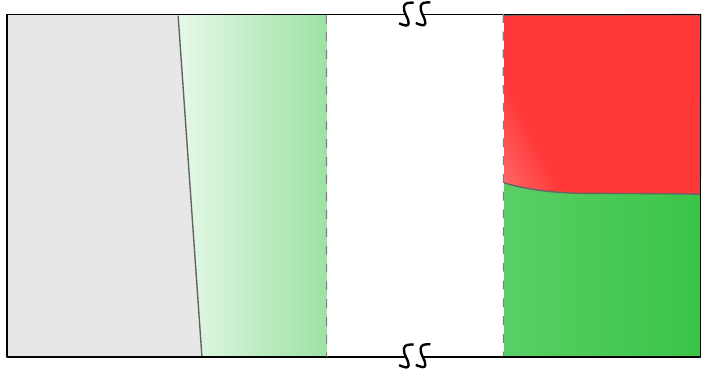}
 \put(-15,47){\textbf{(b)}}
 \put(01,-5){$\leftarrow \, \lambda \gg zJ_\mathrm{SE}$}
 \put(71,-5){$zJ_\mathrm{SE} \gg  \lambda \; \rightarrow$}
 \put(-6,16){\rotatebox{90}{$J_H / U \; \rightarrow$}}
 \put(3,24){van Vleck}
 \put(10,18){PM}
 \put(33,27){AF}
 \put(28.7,21){Triplon}
 \put(31.5,15){BEC}
 \put(76.5,11.5){Spin AF}
 \put(78,35){Spin F}
\end{overpic}
\end{center}
\vspace{0.1cm}

\caption{
\coloronline (a) The single site total angular momentum is zero in both the $jj$ and $LS$ coupling schemes. (b) Schematic phase diagram of the spin-orbital model appearing in \eqref{pertLatticeHamiltonian} pitting spin-orbit coupling $\lambda$ against superexchange $J_\mathrm{SE}$ where $\lambda$ is the spin-orbit coupling energy scale and $J_\mathrm{SE}$ is the superexchange energy scale with $z$ being the coordination number. Starting with a van Vleck phase with no atomic moments at large $\lambda$ we find a triplon condensate at $\veck{k} =\vec\pi$ for all values of the Hund's coupling $J_H/U$. The intermediate regime where $\lambda \approx z J_\mathrm{SE}$ has not been explored. At large $J_\mathrm{SE}$ we obtain effective magnetic Hamiltonians that have isotropic Heisenberg spin interactions (antiferromagnetic for small $J_H/U$ and ferromagnetic for large $J_H/U$) but the orbital interactions are more complex and anisotropic. We expect novel magnetic phases arising from orbital frustration in the intermediate and large $J_\mathrm{SE}/\lambda$ regimes.
\label{AfmFmSchematicPhases}
}
\end{figure}
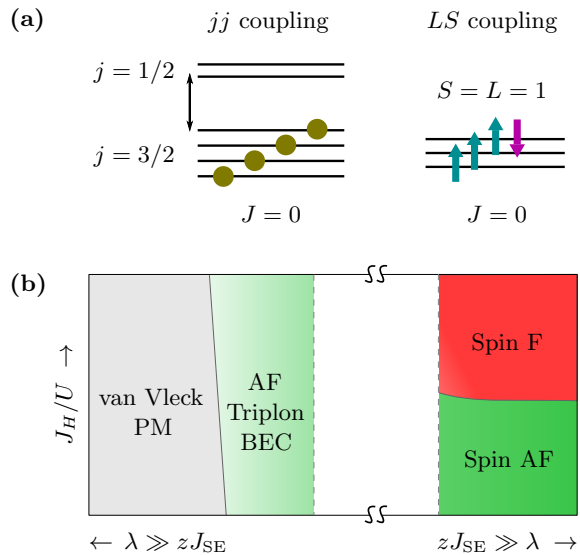

In transition metal oxides with oxygen octahedra, the large crystal field splitting puts $d^4$ ions into the $t_{2g}^4$ electronic configuration.
For materials with strong spin-orbit coupling, the $j=3/2$ band is filled and the $j=1/2$ band is empty leading to the conclusion that weakly correlated $d^4$ materials are non-magnetic.
However when Coulomb interactions are strong, a total spin $S=1$ and orbital angular momentum $L=1$ lead to a total angular momentum $J=0$ on every $d^4$ ion with no magnetism.
Thus both {\it jj} coupling and {\it LS} coupling schemes lead to the same conclusion that a single atom is in a $J=0$ singlet state and therefore non-magnetic as shown in Fig.~\ref{AfmFmSchematicPhases}(a).

We build on previous work by Khaliullin\cite{d4Khaliullin2013} that proposed an ``exciton condensation" mechanism, more accurately a condensation of $J=1$ triplon excitations, to drive the onset of antiferromagnetism in nominally non-magnetic $d^4$ systems and our previous study\cite{Meetei2015} showing that ferromagnetic superexchange interactions caused by strong Hund's coupling can precipitate ferromagnetic coupling.
In this work we start with the atomic multi-orbital Hamiltonian with intra- and inter-orbital Coulomb interactions and spin-orbital coupling specifically for $t_{2g}^4$ systems.
We next allow hopping between atoms and investigate all cases of orbital geometries-- the idealized fully symmetric case when all orbitals participate in hopping, as well as more realistic cases suitable for simple cubic and face-centered cubic lattices.
For each case, we derive the effective spin-orbital superexchange Hamiltonian which competes with spin-orbit coupling to produce strong deviations from the non-magnetic atomic behavior.
These results are obtained both using exact diagonalization on a two-site problem and perturbation theory for the effective magnetic interactions.

Tuning the superexchange interactions $J_\mathrm{SE}$ relative to spin-orbit coupling $\lambda$, we first see the formation of local moments followed by a Bose condensation of weakly interacting $J=1$ triplet excitations, or triplon condensation.
Rather remarkably, regardless of the local spin interactions favoring antiferromagnetic spin superexchange (spin-AF) at small Hund's coupling or ferromagnetic spin superexchange (spin-F) at large Hund's coupling, the $J=1$ triplons condense at the $\vec k=\vec\pi$ point.
This result that the rotationally invariant spin-orbit coupling can effectively flip the sign of superexchange is unusual and unique to spin-orbital coupled systems. 
In the opposite regime where $J_\mathrm{SE}$ dominates, the orbital interactions are frustrated even in the absence of geometric frustration and can potentially lead to orbital liquid phases. Even when $\lambda=0$ and the local spin interactions are simple Heisenberg FM or AFM, the frustrated orbital interactions generate frustration in the spin channel as well, leading to the possibility of ground states with both orbital and spin entanglement on lattices without geometric frustration.
This is summarized schematically in Fig.~\ref{AfmFmSchematicPhases}(b).

The paper is organized in the following way.
In Section~\ref{sectionModel} we introduce the lattice Hamiltonian used as the basis for the rest of the paper which includes electron hopping, atomic spin-orbit coupling, and an effective multi-orbital Coulomb interaction that captures Hund's rules.
The orbital geometries for hopping used throughout the paper include both a highly symmetric toy model to be used as a simplified diagnostic tool as well as two other more realistic cases found in perovskites.

In Section~\ref{sectionExactDiagonalization} we use exact diagonalization to study a two-site specialization of the problem introduced in Section~\ref{sectionModel}.
Ref.~\citenum{IsobeNagaosa2014} has used a similar procedure to study transition metal systems with other electron counts.
Local magnetic moments are absent when spin-orbit coupling is large, as expected in the atomic picture, but electron hopping introduces sizeable moments when $t\sim \lambda$ when two or three orbitals strongly overlap between sites.
Although a single orbital overlap can also promote superexchange which competes with spin-orbit coupling, the number of superexchange pathways is limited and local moments do not form for any reasonable ratio of $t/\lambda$.

In Section~\ref{sectionEffectiveHamiltonian} we derive an effective magnetic Hamiltonian in terms of orbital angular momentum and spin operators using second order perturbation theory.
We check that the spin-orbital superexchange Hamiltonian captures both spin-AF and spin-F interactions between spins depending on the value of Hund's coupling, and the sum of spin-orbit coupling and the spin-orbital superexchange Hamiltonian reproduce the phases found in exact diagonalization of a two-site system.

In Section~\ref{orbitalFrustrationSubsection}, we give a qualitative description of how bond-dependent spin-orbital superexchange results in orbital frustration.
However, finding solutions to orbitally frustrated models can be challenging and is outside the scope of the present paper.

In Section~\ref{NatureOfMagneticPhases} we first review the ``excitonic'' condensation mechanism where the Bose condensation of van Vleck excitations gives magnetism to $d^4$ systems with spin-orbit coupling.
Although the condensation mechanism involves approximations to full spin-orbital models derived in the previous section, it gives valuable insight into the behavior at large spin-orbit coupling.
Regardless of the nature of local interactions, only AF condensates are supported for the models studied, and we give the the critical superexchange required for AF condensation for the three orbital geometries studied.

Section~\ref{MaterialsExperimentsSection} discusses potential materials realizations and experiments beyond those mentioned in the introduction.

\section{Model}
\label{sectionModel}
Our model Hamiltonian for $t_{2g}$ systems is composed of three parts: (i) kinetic part, (ii) Coulomb interaction, and (iii) spin-orbit coupling.
\begin{equation}
H = \sum_{\langle ij \rangle} H_{\mathrm{t}}^{(ij)} + \sum_{i} H_{\mathrm{int}}^{(i)} + \sum_{i} H_{\mathrm{so}}^{(i)}
\label{HexactDiag}
\end{equation}
The general form of the kinetic part
\begin{equation}
H_{\mathrm{t}}^{(ij)} = \sum_{mm'} \sum_{\sigma} t_{m'm}^{(ij)} \, \cre{c}{im'\sigma} \ann{c}{jm\sigma} + \mathrm{h.c.}
\label{Ht}
\end{equation}
is given in terms of matrix elements $t_{m'm}^{(ij)}$ between $t_{2g}$ orbitals $m'$ and $m$ (with values $yz$, $zx$, and $xy$) on sites $i$ and $j$.
The index $\sigma$ is for spin.
We take the on-site Coulomb interaction to be the $t_{2g}$ interaction Hamiltonian \cite{doi:10.1146/annurev-conmatphys-020911-125045}
\begin{equation}
H_{\mathrm{int}}^{(i)} = (U-3J_H)\frac{N_i (N_i-1)}{2} + J_H \left( \tfrac{5}{2} N_i - 2 S_i^2 - \tfrac{1}{2} L_i^2 \right)
\label{Ht2g}
\end{equation}
where $N$ is electron number, $S$ is total spin, and $L$ is total orbital angular momentum.
The on-site intra-orbital Hubbard interaction is characterized $U$ and $J_H$ characterizes the strength of Hund's coupling.
We have chosen to use $J_H$ instead of $J$ to avoid confusion with total angular momentum in the next two sections.
See Appendix A for details.

The atomic spin-orbit coupling has the form
\begin{equation}
H_{\mathrm{so}}^{(i)} = \lambda \sum_{m m'} \sum_{\sigma \sigma'} \cre{c}{im'\sigma'} \left( \veck{l}_{m'm} \cdot \veck{s}_{\sigma' \sigma} \right) \ann{c}{im\sigma}
\label{Hso}
\end{equation}
where $\veck{l}$ is the projection of angular momentum operators to the $t_{2g}$ subspace, $(l_k )_{m'm} = i \epsilon_{k m' m}$ so that $\veck{l} \times \veck{l} = -i \veck{l}$, and $\veck{s}$ is the spin operator with $\veck{s} = \veck{\sigma}/2$ where we have set $\hbar = 1$.
\begin{figure}
\begin{overpic}[width=2.9in]{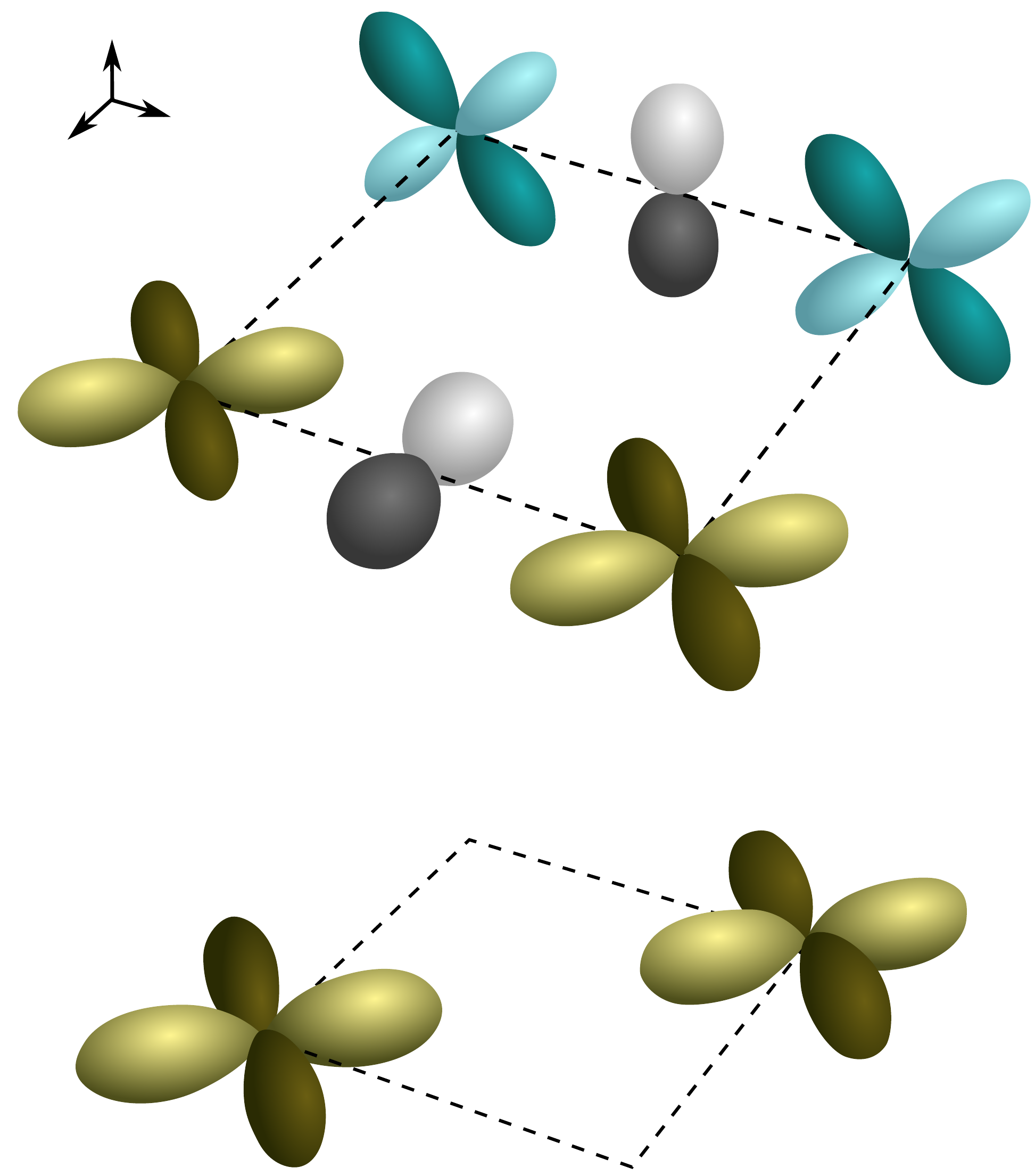}
 \put(3,87){$x$}
 \put(15,89){$y$}
 \put(8.5,97.5){$z$}
 \put(4,75){$d_{xy}$}
 \put(24,97){$d_{yz}$}
 \put(55,64){$d_{xy}$}
 \put(65,90){$d_{yz}$}
 \put(52,94){$p_z$}
 \put(36,70){$p_x$}
 \put(11,23){$d_{xy}$}
 \put(60,32){$d_{xy}$}
 \put(-3,97){\textbf{(a)}}
 \put(-3,35){\textbf{(b)}}
\end{overpic}
\caption{
\coloronline (a) The $N_\mathrm{orb}=2$ model is an approximation of oxygen mediated electron hopping between $t_{2g}$ orbitals in a simple cubic lattice.  Both $d_{xy}$ and $d_{yz}$ orbitals participate in hopping along the $y$ direction. (b) The $N_\mathrm{orb}=1$ model is an approximation of direct hopping between $t_{2g}$ orbitals on the face of a face-centered cubic lattice. The $d_{xy}$ orbitals are most relevant for hopping in the $xy$ plane.  
\label{latticeGeometriesFigure}
}
\end{figure}
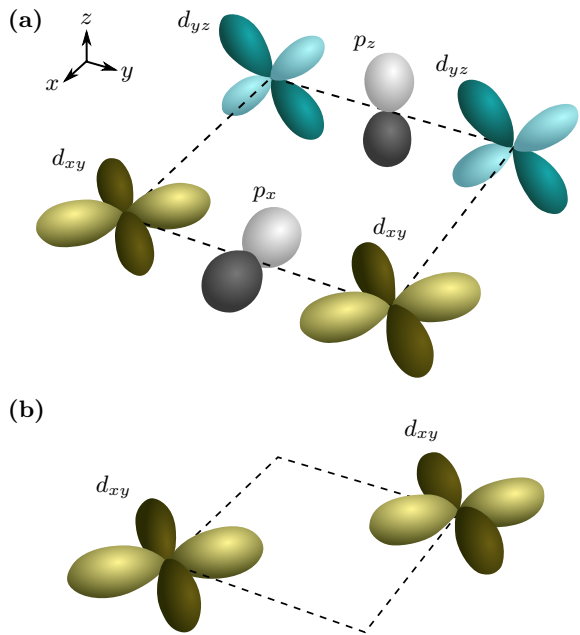

We focus on three special cases of $t_{m'm}^{(ij)}$ which differ by the number of orbitals, $N_\mathrm{orb}$ participating in hopping.
\begin{itemize}[leftmargin=*]
\item $N_\mathrm{orb} = 3$: First we consider the orbitally symmetric case where $t_{m'm}^{(ij)} = t \delta_{m'm}$ and all orbitals participate in hopping.
While this full rotational symmetry is not usually present in material systems, the $N_{\mathrm{orb}} = 3$ case serves as a diagnostic tool where total angular momentum in the system is conserved and correlation functions have rotational symmetry.
\item $N_\mathrm{orb} = 2$: The next case, $t_{m'm}^{(ij)} = t \delta_{m'm} \left( 1 - \delta_{k m} \right)$, uses two orbitals participating in hopping, $N_{\mathrm{orb}}= 2$, while one orbital is blocked.
The blocked orbital $k$ is determined by the direction of the line connecting sites $i$ and $j$.
This situation is commonly found in simple cubic lattices where $t$ comes from oxygen-mediated superexchange. 
See Fig.~\ref{latticeGeometriesFigure}(a).
\item $N_\mathrm{orb} = 1$: The final case, $t_{m'm}^{(ij)} = t \delta_{m'm} \delta_{k m}$, only has one orbital contributing, $N_\mathrm{orb} = 1$, while two orbitals are blocked and approximates the hopping between nearest-neighbors on a face-centered cubic lattice. 
The active orbital $k$ is determined by which plane the sites $i$ and $j$ share.
See Fig.~\ref{latticeGeometriesFigure}(b).
\end{itemize}

\section{Exact diagonalization}
\label{sectionExactDiagonalization}
\begin{figure}
\begin{flushright}
\begin{overpic}[width=8.23cm]{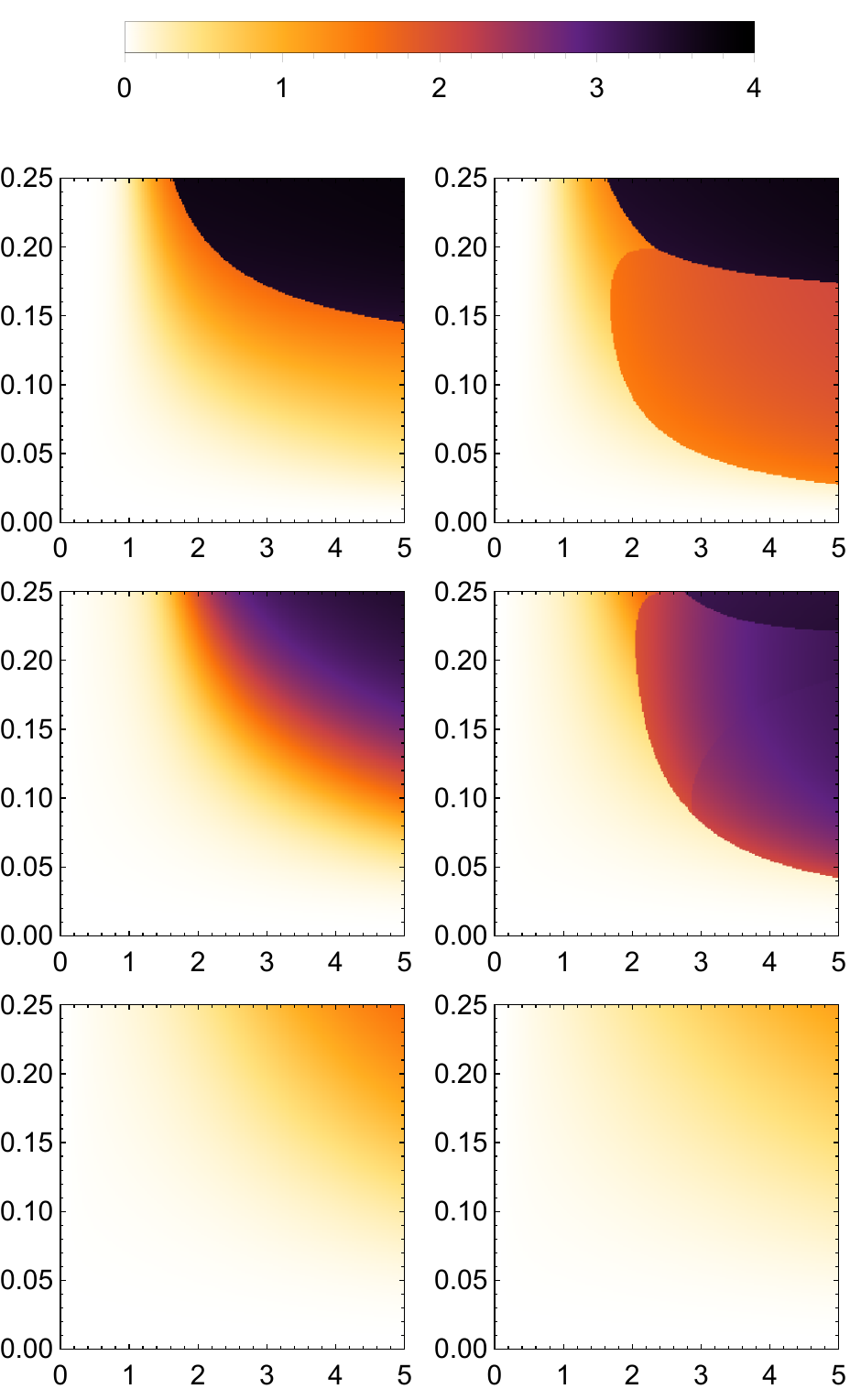}
 \put(5,84.5){\textbf{(a)}}
 \put(36,84.5){\textbf{(b)}}
 \put(5,55){\textbf{(c)}}
 \put(36,55){\textbf{(d)}}
 \put(5,25.5){\textbf{(e)}}
 \put(36,25.5){\textbf{(f)}}
 \put(10,88.7){$J_H / U = 0.1$}
 \put(41,88.7){$J_H / U = 0.2$}
 \put(3.5,96.7){$\langle J_i^2 \rangle$}
 \put(-3.3,43){\rotatebox{90}{$t\,/\,U$}}
 \put(29,-2){$t\,/\,\lambda$}
 \put(5,64){{$N_{\mathrm{orb}} = 3$}}
 \put(36,64){{$N_{\mathrm{orb}} = 3$}}
 \put(5,34.7){{$N_{\mathrm{orb}} = 2$}}
 \put(36,34.7){{$N_{\mathrm{orb}} = 2$}}
 \put(5,5){{$N_{\mathrm{orb}} = 1$}}
 \put(36,5){{$N_{\mathrm{orb}} = 1$}}
\end{overpic}
\hspace{0.00cm}
\end{flushright}
\caption{
\coloronline The Hamiltonian in equation \eqref{HexactDiag} is solved for a two-site system.
The local total angular momentum squared on one site, $\langle J_i^2 \rangle$, is plotted for small and large values of Hund's coupling, $J_H / U = 0.1$ and $J_H / U = 0.2$, for the three types of hopping matrices used in the text.
(a-b) Hopping using $N_{\mathrm{orb}} = 3$ produces sizable local moments.
For small Hund's coupling, the local moment gradually forms as $t$ is turned on.
For large Hund's coupling, there is an abrupt formation of large local moments due to an energy level crossing.
(c-d) Hopping using $N_{\mathrm{orb}} = 2$ produces qualitatively similar behavior to the $N_{\mathrm{orb}} = 3$ case.
(e-f) Hopping using $N_{\mathrm{orb}} = 1$ produces negligible moments.
\label{conservedJphasediagram}
}

\vspace{0.15cm}
\begin{overpic}[width=8.7cm]{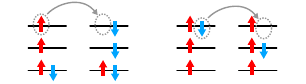}
 \put(0,22){\textbf{(a)}}
 \put(48,22){\textbf{(b)}}
\end{overpic}
\caption{
\coloronline (a) The virtual process leaves the first site in a low spin, $S=\tfrac{1}{2}$, configuration and results in antiferromagnetic superexchange.
(b) The virtual process leaves the first site in a high spin, $S=\tfrac{3}{2}$, configuration and results in ferromagnetic superexchange.
\label{AfmFmMechanism}
}
\end{figure}

Before analyzing the full lattice problem which will require approximations to be made, it is useful to examine exact results for a pair of interacting sites.
We numerically diagonalize \eqref{HexactDiag} for a two-site site system, with site labels $i$ and $j$, to extract the magnetic interactions in the Mott limit.
We choose the blocked orbital $k$ to be the $xy$ orbital for the $N_\mathrm{orb}=2$ and $N_\mathrm{orb}=1$ models.
Fig.~\ref{conservedJphasediagram} gives ground state values of the square of the local total angular momentum, $\langle J_i^2 \rangle$, for the two-site specialization of \eqref{HexactDiag}.
For all three types of hopping matrices, small $t$ compared to $\lambda$ give negligible local moments since spin-orbit coupling keeps each site in a nonmagnetic $J_i = 0$ spin-orbital singlet.
For larger values of $t$, local moments may form from the tendency of superexchange to cause spin and orbital ordering which is incompatible with local spin-orbital singlet behavior on each site.
For both $N_{\mathrm{orb}} = 3$ and $N_{\mathrm{orb}} = 2$, this effect is pronounced and requires $t/\lambda \approx 2$ at the two-site level.
In a lattice, this critical ratio will be reduced due to presence of many neighboring sites contributing to superexchange, hence a smaller hopping $t$ is able to destabilize the atomic singlet.
For $N_{\mathrm{orb}} = 1$, the effect is much less pronounced since the number of superexchange paths is limited.

When a single orbital is active, the results do not sensitively depend on $J_H / U$, however, the presence of strong Hund's coupling results in qualitatively different behavior for the $N_{\mathrm{orb}} = 3$ and $N_{\mathrm{orb}} = 2$ models.
We expect that antiferromagnetic superexchange between spins (spin-AF) is responsible for moment formation and can qualitatively be understood in the following way.
Each site has a local total spin $S_i =1$ and local orbital angular momentum $L_i =1$ ($^3 P$ configuration) from each $t_{2g}$ orbital being at least singly occupied with one of the three orbitals doubly occupied.
To maximize the number of superexchange paths, the orbitals participating in antiferromagnetic superexchange should be singly occupied.
This means the doubly occupied orbitals try to match up between neighboring sites, see Fig.~\ref{AfmFmMechanism}(a).
When each pair of singly occupied orbitals between sites is in a spin singlet, the two-site system is a spin singlet, but the orbitals are in a ferro-orbital state.
The combined spin-AF and orbital-F interactions are responsible for moment formation in Figs.~\ref{conservedJphasediagram}(a) and \ref{conservedJphasediagram}(c).

Large values of Hund's coupling can produce a different ground state via a level crossing at the sharp boundaries in Figs.~\ref{conservedJphasediagram}(b) and \ref{conservedJphasediagram}(d) which are not present in Figs.~\ref{conservedJphasediagram}(a) and \ref{conservedJphasediagram}(c).
This behavior can be understood by examining the effect of Hund's coupling on the $d^3 d^5$ states during the virtual $d^4d^4\rightarrow d^5d^3 \rightarrow d^4d^4$ process.
The intermediate $d^3$ may have either a maximized spin state $S_i=\tfrac{3}{2}$ or a minimized spin state $S_i=\tfrac{1}{2}$, and large values of Hund's coupling make the intermediate $S_i =\tfrac{3}{2}$ states energetically more favorable.
Moving an electron off a doubly-occupied orbital leaves the ion in an energetically favorable $S_i =\tfrac{3}{2}$ configuration.
For example, see Fig.~\ref{AfmFmMechanism}(b).
Since this requires electrons to move onto single-occupancy orbitals on the other site, the Goodenough-Kanamori-Anderson rules\cite{goodenough1963,Kanamori1959,Anderson1959} indicate the the spin interactions are ferromagnetic (spin-F) but the orbitals are in a singlet state (orbital-AF). Here too, as in the small Hund's coupling case, on a lattice when an electron hops along different directions,
the doubly occupied orbitals become bond-dependent and lead to 
anisotropic interactions.

\section{Effective Magnetic Hamiltonian}
\label{sectionEffectiveHamiltonian}
We now begin our analysis of the full lattice problem in Eq.~\eqref{HexactDiag} by calculating the effective spin-orbital lattice model for each of the three $N_\mathrm{orb}$ cases.
Only the main results are presented here; the details of the calculation are presented in Appendix \ref{effectiveHamiltonianAppendix}.

To understand the superexchange mechanisms in the three different $N_\mathrm{orb}$ models and how they compete with spin-orbit coupling, we derive an effective magnetic spin-orbital Hamiltonian\cite{KugelKhomskii} within the local $^3 P$ space (spectroscopic notation $^{2S+1} L_J$) on each site.
This effective Hamiltonian is written as the sum of spin-orbit and superexchange terms.
\begin{equation}
H_{\mathrm{eff}} = \sum_i H_{\mathrm{SOC}}^{(i)} + \sum_{\langle ij \rangle} H_{\mathrm{SE}}^{(ij)}
\label{pertLatticeHamiltonian}
\end{equation}
The first order spin-orbit correction $H_\mathrm{SOC}^{(i)} = \tfrac{\lambda}{2} \veck{L}_{i} \cdot \veck{S}_{i}$ is qualitatively correct, but we give the second order effective spin-orbit interaction within the local $^3 P$ space to numerically match the energies from exact diagonalization.
\begin{equation}
H_{\mathrm{SOC}}^{(i)} = \frac{\lambda}{2} \left( 1 - \frac{1}{4} \frac{\lambda}{J_H} \right) \veck{L}_i \cdot \veck{S}_i - \frac{7}{40} \frac{\lambda^2}{J_H} \left( \veck{L}_i \cdot \veck{S}_i \right)^2
\label{effectiveSOCHamiltonian}
\end{equation}
The spin-orbital superexchange Hamiltonian, $H_{\mathrm{SE}}$, is constructed using three different virtual exchange processes each defined by the energy values of intermediate multiplets.\cite{Ishihara1996,Ishihara1997,Oles1999}
In the present case, the $d^3$ electron configuration in the virtual process $d^4 d^4 \rightarrow d^3 d^5 \rightarrow d^4 d^4$ is used to label these superexchange pathways.\cite{GiniyatOles2001,Perkins2002}
Each pathway yields a superexchange term which is the product of a spin interaction and a $t_{m'm}^{(ij)}$-dependent orbital interaction $\mathcal{O}_{ij}$.
Since $^4 S$, $^2 D$, and $^2 P$ label the intermediate $d^3$ configurations, we arrive at the three corresponding superexchange terms.
\begin{equation}
\begin{split}
H_{\mathrm{SE}}^{(ij)} =
 - \tfrac{t^2}{U - 3J_H} \left(2 + \veck{S}_i \cdot \veck{S}_j \right) \mathcal{O}_{ij}^{S} \\
 - \tfrac{t^2}{U} \left(1 - \veck{S}_i \cdot \veck{S}_j \right) \mathcal{O}_{ij}^{D} \\
 - \tfrac{t^2}{U + 2J_H} \left(1 - \veck{S}_i \cdot \veck{S}_j \right) \mathcal{O}_{ij}^{P}
\end{split}
\label{effectiveSEHamiltonian}
\end{equation}
The first pathway, $\mathcal{O}_{ij}^{S}$ corresponding to the $^4 S$ state, has the lowest energy of all the intermediate states since maximizing the spin of the $d^3$ configuration minimizes the total energy.
We see that maximizing the total spin favors spin-F behavior as in Fig.~\ref{AfmFmMechanism}(b).
The other two pathways, $\mathcal{O}_{ij}^{D}$ and $\mathcal{O}_{ij}^{P}$ corresponding to $^2 D$ and $^2 P$, minimize the total spin with $S_i = 1/2$ in the $d^3$ configuration and will favor spin-AF as in Fig.~\ref{AfmFmMechanism}(a).

The spin-F and spin-AF behaviors may be verified by observing that both $2+\veck{S}_i \cdot \veck{S}_j$ and $1-\veck{S}_i \cdot \veck{S}_j$ are non-negative.
Then the energy due to each pathway may be minimized by simultaneously maximizing the spin part and the orbital part separately.
Since each hopping matrix $t_{m'm}$ uniquely determines the resulting orbital interactions $\mathcal{O}_{ij}$, we will compute these orbital interactions explicitly for previous three choices of $t_{m'm}$ ($N_\mathrm{orb} = 3,2,1$).
We will find that $^2 {D}$ and $^2 {P}$ pathways will together dominate over the $^4 S$ pathway when $J_H / U$ is small, however, this can change at larger values of $J_H / U$.

The spin-orbital models we calculate here are similar to those of $d^2$ systems.\cite{GiniyatOles2001,Perkins2002}
This follows from the fact that a $(t_{2g})^4$ electronic system is the particle hole conjugate to a $(t_{2g})^2$ hole system.
Formally every $(t_{2g})^2$ spin-orbital model may be transformed into a $(t_{2g})^4$ spin-orbital model so long as (a) the crystal field splitting is large enough to prevent high spin configurations from becoming energetically relevant and (b) the fundamental parameters $\lambda$ and $t_{m'm}$ are negated.

Before proceeding with the explicit calculations for $\mathcal{O}_{ij}$, we note the intimate connection between the type of spin state favored (ie. spin-AF or spin-F) and the orbital state pictured in Fig.~\ref{AfmFmMechanism} is now mathematically depicted in \eqref{effectiveSEHamiltonian}.
While each pathway contributes a spin-orbital term which is the product of spin and an orbital term, the sum of the three pathways cannot generally be factored in this way.
The consequence is that even \textit{without} the spin-orbit interaction, the spins and orbitals are \textit{not} independent on a site in the lattice\cite{PhysRevB.75.195113} even though they \textit{are} independent at the atomic level.
This can have important consequences on the types of ordering allowed in lattices when the orbital part becomes frustrated due to orbital geometry even on geometrically unfrustrated lattices.\cite{Oles2012}

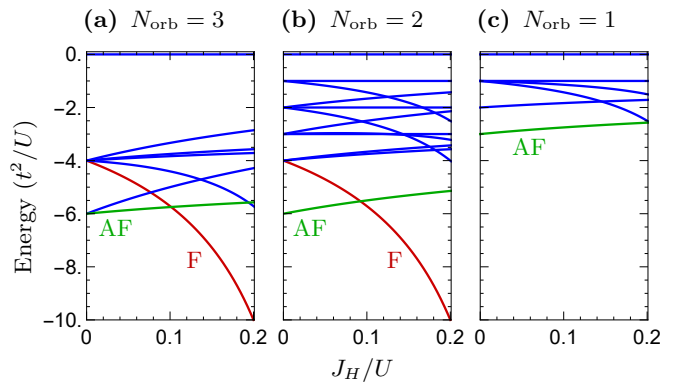
\begin{figure}
\vspace{0.45cm}
\begin{flushright}
\begin{overpic}[width=8.33cm]{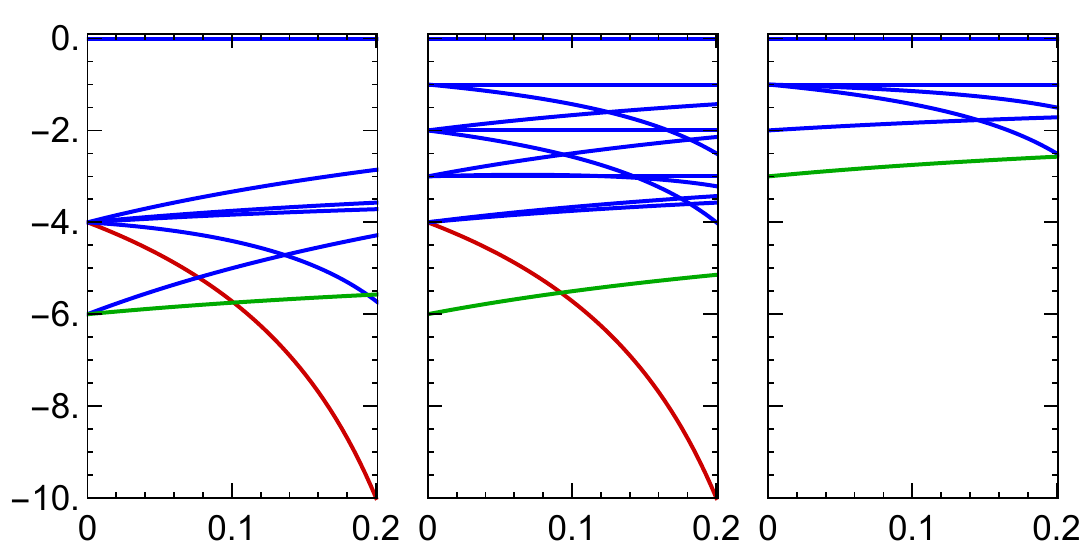}
 \put(7.5,52.5){\textbf{(a)}\hspace{0.05cm} $N_\mathrm{orb} = 3$}
 \put(39.0,52.5){\textbf{(b)}\hspace{0.05cm} $N_\mathrm{orb} = 2$}
 \put(70.3,52.5){\textbf{(c)}\hspace{0.05cm} $N_\mathrm{orb} = 1$}
 
 \put(-4,13){\rotatebox{90}{Energy $(t^2/U)$}}
 \put(47,-3.3){$J_H / U$}
 
 \put(24,14){\color{myred2}F}
 \put(10,19){\color{mygreen2}AF}
 
 \put(56,14){\color{myred2}F}
 \put(41,19){\color{mygreen2}AF}
 
 \put(76,32){\color{mygreen2}AF}
\end{overpic}
\end{flushright}
\caption{
\coloronline 
Energy eigenvalues of the two-site superexchange Hamiltonian \eqref{effectiveSEHamiltonian} are plotted for (a) $N_\mathrm{orb} = 3$ using \eqref{orbSymmChannelAll}, (b) $N_\mathrm{orb} = 2$ using \eqref{orb2ChannelAll}, and (c) $N_\mathrm{orb} = 1$ using \eqref{orb1ChannelAll}.
In addition to a spin-AF ground state, a spin-F ground state can be favored when Hund's coupling is large in both the $N_\mathrm{orb} = 3$ and $N_\mathrm{orb} = 2$ models.
\label{energySpectra}
}
\end{figure}

\subsection{$N_\mathrm{orb} = 3$}
\label{3activeOrbitalSection}
For the orbitally symmetric model, $t_{m'm} = t \delta_{m'm}$, the total orbital angular momentum, $L$, is conserved.
We obtain the effective superexchange terms for this interaction term below.
\begin{subequations}
\label{orbSymmChannelAll}
\begin{align}
& \mathcal{O}_{ij}^{S} =  \tfrac{4}{3} - \tfrac{2}{3} \veck{L}_i \cdot \veck{L}_j - \tfrac{2}{3} (\veck{L}_i \cdot \veck{L}_j )^2  
\label{orbSymmChannelA}
\\
& \mathcal{O}_{ij}^{D} = \tfrac{4}{3} + \tfrac{1}{3} \veck{L}_i \cdot \veck{L}_j - \tfrac{1}{6} (\veck{L}_i \cdot \veck{L}_j )^2  
\label{orbSymmChannelB}
\\
& \mathcal{O}_{ij}^{P} = \tfrac{1}{2} (\veck{L}_i \cdot \veck{L}_j )^2 
\label{orbSymmChannelC}
\end{align}
\end{subequations}
To understand these results, we exactly diagonalize the effective superexchange Hamiltonian in the context of a two-site system.
It is useful to rewrite each pathway in terms of projection operators, $\mathbb{P}$, to a particular subspace of total angular momentum $L = 0,1,2$.
\begin{subequations}
\begin{align}
& \mathcal{O}_{ij}^{S} =  \tfrac{4}{3} \mathbb{P}(L=1)
\\
& \mathcal{O}_{ij}^{D} = \tfrac{5}{6} \mathbb{P}(L=1) + \tfrac{3}{2} \mathbb{P}(L=2)
\\
& \mathcal{O}_{ij}^{P} = 2 \mathbb{P}(L=0) + \tfrac{1}{2} \mathbb{P}(L=1) + \tfrac{1}{2} \mathbb{P}(L=2)
\end{align}
\end{subequations}
Fig.~\ref{energySpectra}(a) shows the energy levels of this superexchange Hamiltonian for different $J_H / U$ for a two-site problem.
Owing to the fact that \eqref{orbSymmChannelA} can be written as the projection to a total angular momentum $L=1$ shared along a bond (up to a factor of $4/3$), the ground states of the $^4 S$ pathway in \eqref{effectiveSEHamiltonian} have total $L=1$ and total $S=2$ shared between the two sites.
Spin-orbit coupling will split these states and make local interactions favor a total $J=1$ shared along a bond.
For large values of $J_H / U$, the $^4 S$ pathway will dominate and the non-zero angular momentum shared between sites gives an effective Curie moment to the two-site system.
Small values of $J_H / U$ will be dominated by the $^2 D$ and $^2 P$ pathways which favor total $L=2$ and $S=0$ in opposition to $^4 S$.
The critical value of Hund's coupling where the $S = 2$ quintet overtakes the $S = 0$ singlet, as seen in Fig.~\ref{energySpectra}(a), can be computed analytically as $J_H / U = \tfrac{1}{54} (\sqrt{505} - 17) \approx 0.1$.

\begin{figure}
\vspace{0.4cm}
\begin{center}
\begin{overpic}[width=3.4in]{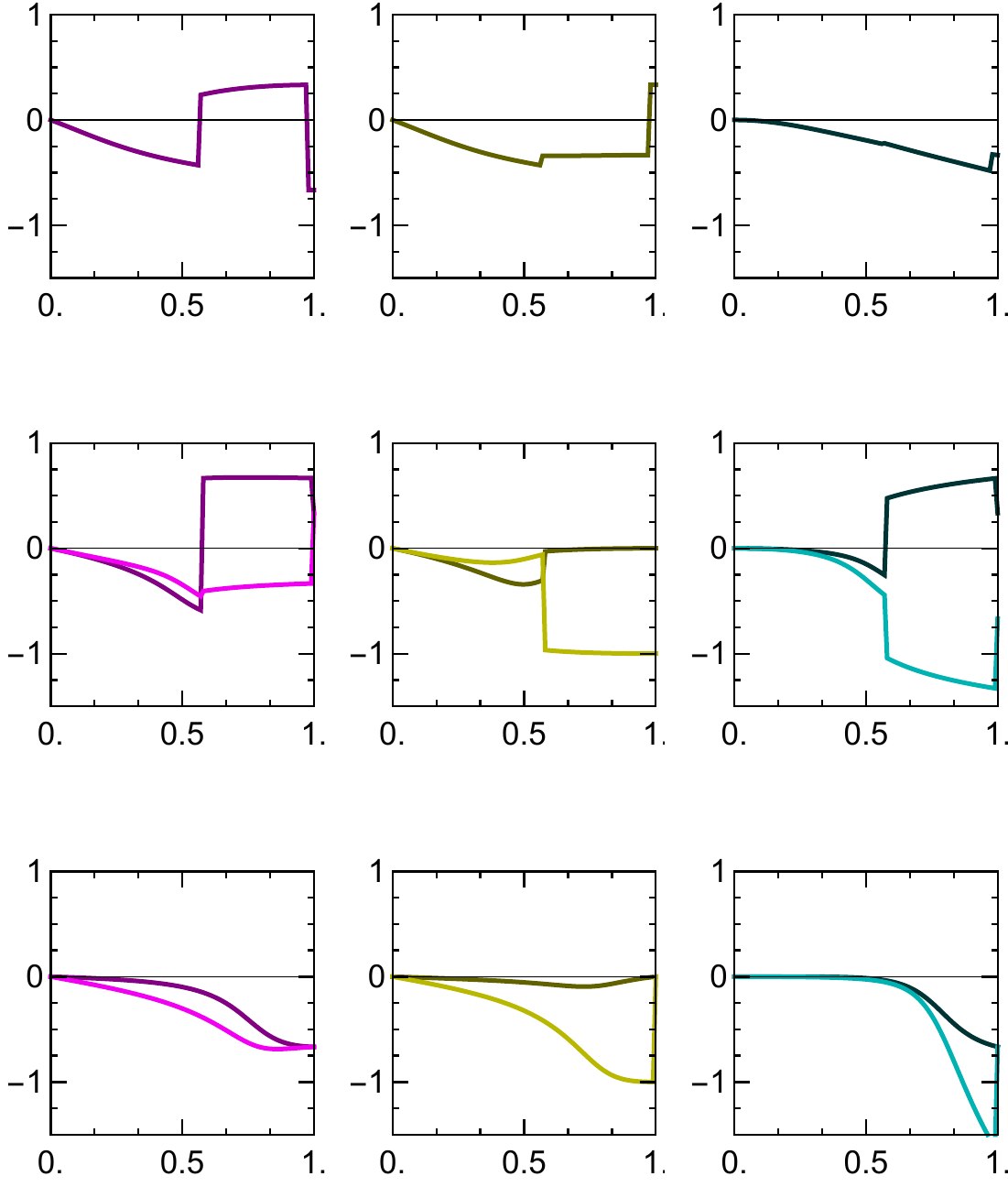}
 \put(38.5,-2.5){$\theta / (\pi / 2)$}
 \put(38.5,34){$\theta / (\pi / 2)$}
 \put(38.5,70){$\theta / (\pi / 2)$}

 \put(18,47){$S_i^z S_j^z$}
 \put(47,46.5){$L_i^z L_j^z$}
 \put(76,45){$J_i^z J_j^z$}
 
 \put(18,56){$S_i^x S_j^x$}
 \put(47,55.5){$L_i^x L_j^x$}
 \put(76,55){$J_i^x J_j^x$}

 \put(12,10){$S_i^z S_j^z$}
 \put(40,10.5){$L_i^z L_j^z$}
 \put(72,10){$J_i^z J_j^z$}
 
 \put(12,19){$S_i^x S_j^x$}
 \put(40,19){$L_i^x L_j^x$}
 \put(76,18.5){$J_i^x J_j^x$}
 
 \put(13,82){$S_i^x S_j^x$}
 \put(42,82){$L_i^x L_j^x$}
 \put(71,84){$J_i^x J_j^x$}
 
 \put(-0.3,101){\textbf{(a)} $N_\mathrm{orb} = 3$}
 \put(-0.3,65){\textbf{(b)} $N_\mathrm{orb} = 2$}
 \put(-0.3,29){\textbf{(c)} $N_\mathrm{orb} = 1$}
\end{overpic}
\end{center}
\caption{
\coloronline
Expectation values of different angular momentum correlators are plotted for the two-site effective Hamiltonian in \eqref{pertLatticeHamiltonian} using the three different $N_\mathrm{orb}$ models with the parameterization $\lambda = \cos \theta$, $t^2/U = \sin \theta$, $J_H / U = 0.1$ and $\lambda / J_H = 1$.
The $N_\mathrm{orb} = 3$ model features full rotational symmetry while the $N_\mathrm{orb} = 2$ and $N_\mathrm{orb} = 1$ models only have one axis of rotational symmetry to make the $z$ correlators different than the $x$ and $y$ correlators.
The effect of increasing $J_H / U$ is to push the crossing point from spin-AF to spin-F behavior further left in these plots.
\label{cuts_figure_2}
}
\end{figure}
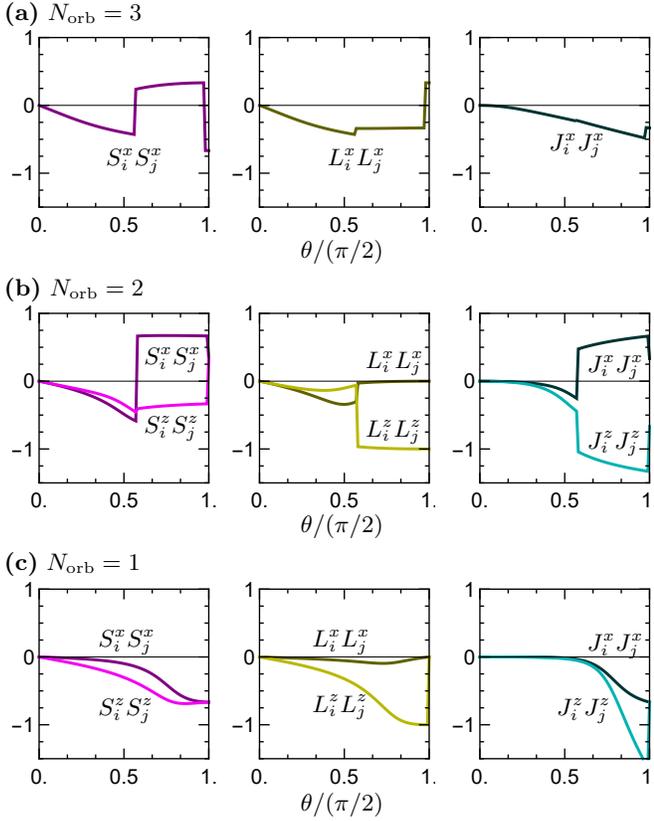

\subsection{$N_\mathrm{orb} = 2$}
The effective superexchange interaction for the two orbital model, $t_{m'm} = t \delta_{m'm} \left( 1 - \delta_{\mathrm{xy},m} \right)$, can be expressed with operators acting on the two active orbitals.
Let $(\veck{\tau}_i,\tau_i^0)$ be the 3+1 Pauli matrices for the $L_i^z = \pm 1$ subspace corresponding to the two active orbitals.
For convenience, we define the permutation operator on the two active orbitals as $P_{ij} = \tfrac{1}{2} ( \veck{\tau}_i \cdot \veck{\tau}_j + \tau _i^0\tau _j^0)$.
Then the orbital part of the superexchange Hamiltonian can be expressed in the following way.
\begin{subequations}
\label{orb2ChannelAll}
\begin{align}
& \mathcal{O}_{ij}^{S} = - \tfrac{2}{3} P_{ij} + \tfrac{1}{3} ( \tau _i^0+\tau_j^0)  
\\
& \mathcal{O}_{ij}^{D} =  1 - \tfrac{1}{6} P_{ij} -\tfrac{1}{6} (\tau _i^0+\tau_j^0)  + \tfrac{1}{2} \tau_i^z\tau _j^z
\\
& \mathcal{O}_{ij}^{P} = 1 + \tfrac{1}{2} P_{ij} - \tfrac{1}{2} (\tau _i^0+\tau_j^0) - \tfrac{1}{2} \tau_i^z\tau _j^z 
\end{align}
\end{subequations}
When $J_H / U \rightarrow 0$, we recover the $d^4$ spin-orbital superexchange Hamiltonian used in Ref.\citenum{d4Khaliullin2013}.
This limit ignores the F spin interactions induced by Hund's coupling.\cite{d4Khaliullin2013}
The above equations for the orbital part combined with both spin-AF and spin-F spin components from \eqref{effectiveSEHamiltonian} give the complete spin-orbital interactions for the 2-orbital model.

It is also worth noting that when the $^2 D$ and $^2 P$ intermediate states are taken to have the same coefficients, rotational invariance within the active orbital subspace can be restored.
Since the $^2 D$ and $^2 P$ pathways have the same $1 - \veck{S}_i \cdot \veck{S}_j$ spin part, these two pathways may be easily combined
\begin{equation}
\mathcal{O}_{ij}^D + \mathcal{O}_{ij}^P = 2 + \tfrac{1}{3} P_{ij} - \tfrac{2}{3} (\tau _i^0+\tau_j^0)
\label{twoactiveCombinedDP}
\end{equation}
so that the $\tau_i^z \tau_j^z$ Ising anisotropy has been eliminated.
This allows us to draw a parallel between the $N_\mathrm{orb} = 3$ and $N_\mathrm{orb} = 2$ models.
In the $N_\mathrm{orb} = 3$ model, the $S=0$ state (spin-AF) was a maximized $L=2$ (orbital-F).
In the $N_\mathrm{orb} = 2$ model, the $S=0$ state is $L_i^z = L_j^z = 0$ as seen in \eqref{twoactiveCombinedDP} since $\mathcal{O}_{ij}$ is to be maximized so that \eqref{effectiveSEHamiltonian} is minimized.
Graphically this is shown in Fig.~\ref{AfmFmMechanism}(a).
This tendency for spin AF to be accompanied by aligned orbitals is common in spin-orbital models.
Similarly, spin F tends to be accompanied by off-alignment of the orbitals as in Fig.~\ref{AfmFmMechanism}(b).

Returning to the full $N_\mathrm{orb} = 2$ case where $\mathcal{O}_{ij}^D$ and $\mathcal{O}_{ij}^P$ are not combined, we diagonalize effective superexchange Hamiltonian for a two-site system.
For a critical value of Hund's coupling, $J_H / U = \tfrac{1}{9} (\sqrt{34}-5) \approx 0.09$, the $S=2$ quintet overtakes the $S=0$ singlet as shown in Fig. \ref{energySpectra}(a).

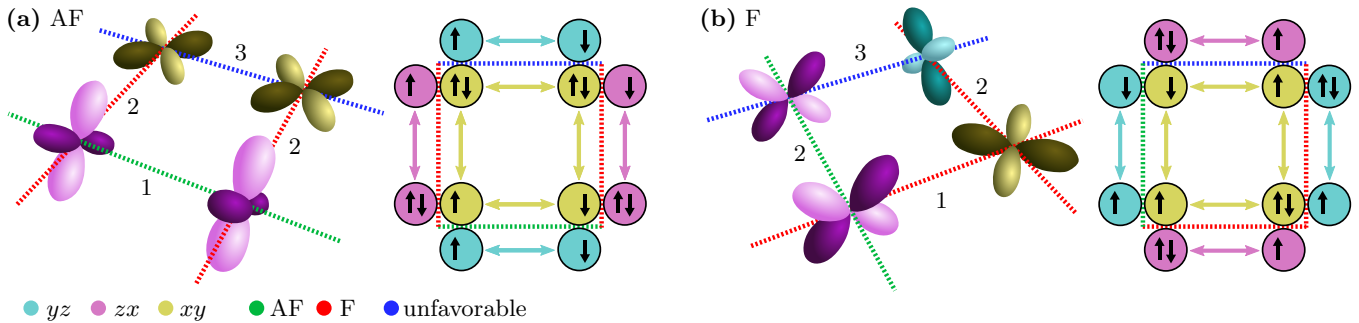
\begin{figure*}
\vspace{0.1cm}
\begin{center}
\begin{overpic}[width=7.05in]{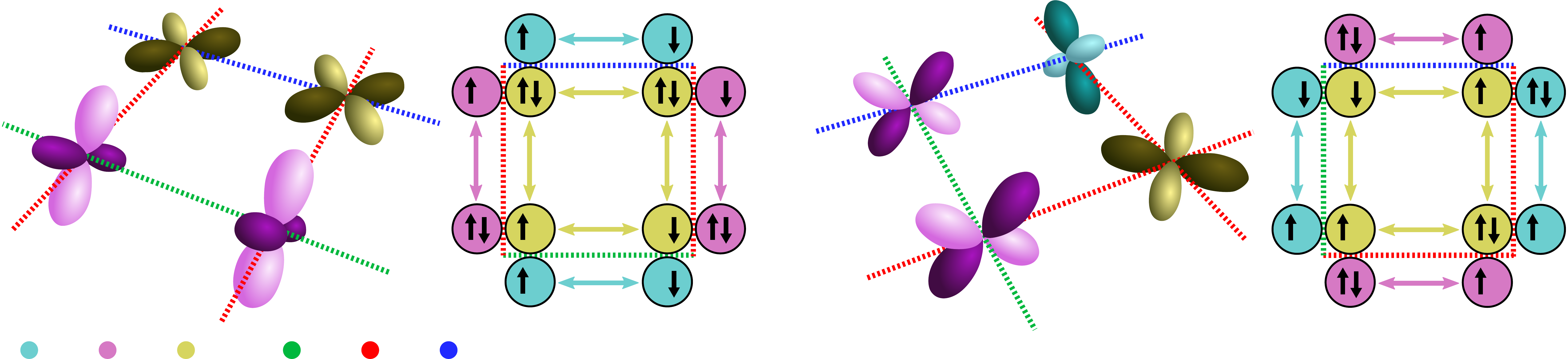}
\put(0,21.5){\textbf{(a)} AF}
 \put(10,9){1}
 \put(9,15){2}
 \put(21,12){2}
 \put(17,19){3}
\put(51.5,21.5){\textbf{(b)} F}
 \put(69,8){1}
 \put(58.5,11.5){2}
 \put(72,16.5){2}
 \put(63,19){3}
 \put(3,0){$yz$}
 \put(8,0){$zx$}
 \put(12.9,0){$xy$}
 \put(19.6,0){AF}
 \put(24.8,0){F}
 \put(29.5,0){unfavorable}
\end{overpic}
\vspace{-0.5cm}
\end{center}
\caption{
\coloronline Orbital frustration is graphically illustrated for the $N_\mathrm{orb} = 2$ model.
The orbitals shown on the vertices of the plaquette are the doubly occupied orbital on each site in a square lattice.
Once the first bond, labeled as 1, is chosen to be of a particular type, either (a) AF or (b) F, the next bonds, labeled as 2, are immediately fixed by this choice.
The result is that the last bond on the plaquette, labeled as 3, then takes a configuration which is neither the most energetically favorable AF bond nor the most energetically favorable F bond.
\label{orbital_frustration_AF_F}
}
\end{figure*}

\subsection{$N_\mathrm{orb} = 1$}
To complete the discussion, we calculate the effective superexchange Hamiltonian for a single orbital hopping model $t_{m'm} = t \delta_{m'm} \delta_{\mathrm{xy},m}$.
\begin{subequations}
\label{orb1ChannelAll}
\begin{align}
& \mathcal{O}_{ij}^{S}  = \tfrac{1}{3} ( L_{i,z}^2 + L_{j,z}^2 ) - \tfrac{2}{3} L_{i,z}^2 L_{j,z}^2 
\\
& \mathcal{O}_{ij}^{D}  = \tfrac{1}{3} ( L_{i,z}^2 + L_{j,z}^2 ) -\tfrac{1}{6} L_{i,z}^2 L_{j,z}^2 
\\
& \mathcal{O}_{ij}^{P}  = \tfrac{1}{2} L_{i,z}^2 L_{j,z}^2 
\end{align}
\end{subequations}
The $^4 S$ pathway is only active when one site is in a $L_{i,z} = \pm 1$ state and the other site is in a $L_{i,z} = 0$ state reflecting that one hole needs to be shared between the sites to allow F.
By combining the $^2 D$ and $^2 P$ pathways as before,
\begin{equation}
\mathcal{O}_{ij}^{D} + \mathcal{O}_{ij}^{P} = \tfrac{1}{3} ( L_{i,z}^2 + L_{j,z}^2 ) + \tfrac{1}{3} L_{i,z}^2 L_{j,z}^2 
\end{equation}
we see AF behavior is maximized when both sites are in the $L_{i,z} = \pm 1$ state reflecting that the $xy$ orbitals participating in superexchange should be singly occupied, and the other two may be doubly occupied.
After diagonalizing the full two-site Hamiltonian in Fig.~\ref{energySpectra}(c), we find this case is qualitatively different from the previous two cases in that F interactions are not supported for any reasonable value of $J_H / U$.
Fig.~\ref{AfmFmMechanism}(b) shows the physical mechanism for F requires active orbitals on opposing sites to share a single hole.
While spin-F states are supported by fewer superexchange paths compared to their spin-AF counterparts, there were many paths for spin-F states to reduce their energy in both the $N_\mathrm{orb} = 3$ and $N_\mathrm{orb} = 2$ models so that Hund's coupling could still tip the balance in favor of spin-F.
However, for $N_\mathrm{orb} = 1$ model, the energy of a spin-F state can only be reduced by a single factor of $-t^2/U$ per site, and favorable spin-F interactions require more than one orbital to be energetically favorable.

\section {Orbital Frustration}
\label{orbitalFrustrationSubsection}

While the orbitally symmetric $N_\mathrm{orb} = 3$ model features  rotational symmetry, the $N_\mathrm{orb} = 2$ and $N_\mathrm{orb} = 1$ models do not due to their orbital geometries.
The $N_\mathrm{orb} = 1$ model requires the single active orbital along a bond to be singly occupied so that AF spin superexchange interactions can minimize the energy.
In $d^4$ configurations, only two of the three orbitals can be singly occupied while one orbital must be doubly occupied.
Since the doubly occupied orbital cannot participate in AF spin superexchange, one third of the bonds must be unsatisfied.

The $N_\mathrm{orb} = 2$ model extends this concept with the possibility for two different low energy states depending on the value of Hund's coupling.
Fig.~\ref{AfmFmMechanism}(a) shows that when two orbitals are active, an AF spin interaction favors double occupancy on the inactive orbital.
An AF bond in the $N_\mathrm{orb} = 2$ model then favors the orbitals perpendicular to the bond direction ($xy$ doubly occupied along a $z$-direction bond) to be doubly occupied.
Bond 1 in Fig.~\ref{orbital_frustration_AF_F}(a) is an example of such an AF bond (green).
Choosing those two doubly occupied orbitals shown in the figure immediately restricts on other bonds emanating from these two sites.
Since the doubly occupied orbitals are not perpendicular to the bonds labeled 2, a different interaction must be favored along the bonds labeled 2.
The next most energetically favorable interaction is the F bond (red) shown in Fig.~\ref{AfmFmMechanism}(b).
This places the double occupancies on the other two orbitals and requires the doubly occupied orbitals on each site to be opposite (ie. $xz$-$yz$ along $z$-direction).
However this leaves the final bond labeled 3 (blue) matching neither the criteria for the lowest energy AF or lowest energy F bond and instead takes an energetically unfavorable AF configuration.
Similarly, starting with an F bond in Fig.~\ref{orbital_frustration_AF_F}(b) as the most energetically favorable results in the same conclusion.
The orbital degrees of freedom then require one of the four bonds on a plaquette to take a high energy configuration in both scenarios.
Regardless of the value of $J_H / U$, the $N_\mathrm{orb} = 2$ model again naturally yields frustration due to the orbital degrees of freedom even on nominally unfrustrated lattices.\cite{Oles2012}
When $\lambda \ll z J_\mathrm{SE}$ where $z$ is the coordination number, these orbital effects are very strong and, in the absence of large octahedral distortions, may lead to orbital liquid states and perhaps highly entangled spin-orbital phases of matter due to quantum fluctuations.

\section{Triplon Condensation}
\label{NatureOfMagneticPhases}
Here we study the case where spin-orbit coupling $\lambda$ is significantly larger than superexchange $J_\mathrm{SE}$, particularly relevant to $5d^4$ materials.
For $J_\mathrm{SE}=0$ the ground state is the product of $J_i=0$ singlets and therefore non-magnetic.
With increasing $J_\mathrm{SE}/\lambda$, a local moment starts to form continuously though long range magnetic order sets in at a finite value of $J_\mathrm{SE}/\lambda$.
This ordered region can be described as triplon condensation of weakly interacting $J=1$ excitations that evolve to a strongly interacting regime.
In this section, we give a detailed introduction to the mechanism of triplon condensation and then apply the formalism to the three $N_\mathrm{orb}$ models considered.

\subsection{Overview of the Mechanism}
\label{sectionTriplonCondensation}
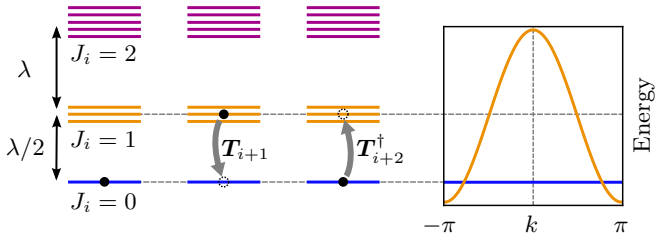
\begin{figure}
\vspace{0.3cm}

\begin{overpic}[width=3.45in,,trim={-0.1cm 0 0 0},clip]{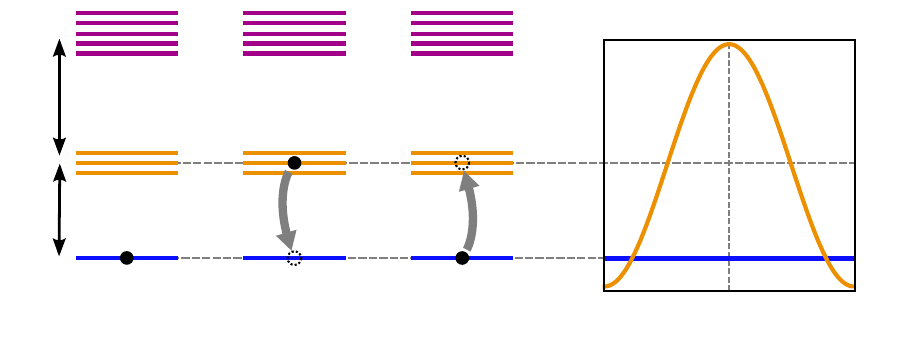}
 \put(-0.2,13.7){$\lambda / 2$}
 \put(1.5,25.5){$\lambda$}
 \put(9.5,5.5){$J_i = 0$}
 \put(9.5,15){$J_i = 1$}
 \put(9.5,28){$J_i = 2$}
 \put(32.5,13.5){$\veck{T}_{i+1}$}
 \put(53,13.5){$\veck{T}_{i+2}^\dagger$}
 \put(78.5,2){$k$}
 \put(63,2){$-\pi$}
 \put(92,2){$\pi$}
 \put(95,14){\rotatebox{90}{Energy}}
\end{overpic}
\caption{
\coloronline The triplon condensation mechanism is graphically illustrated. When there exists a triplon excitation on a site, superexchange can move the excitation to neighboring sites. This effective hopping causes the triplon's energy to disperse in $k$-space. When superexchange becomes large enough, condensation of triplon excitations occurs as the bottom of the triplon band becomes lower in energy than the original $J_i=0$ level. 
\label{CondensationMechanismFigure}
}
\end{figure}

With zero superexchange, the energy cost to make a $J_i = 1$ triplon excitation is $\lambda / 2$.
It was shown previously,\cite{d4Khaliullin2013} using the bond operator formalism,\cite{SachdevBondOperatorPRB1990} that for spin-AF superexchange interactions that are substantially weaker than spin-orbit coupling, the superexchange interactions allow these triplet excitations to propagate from site to site and disperse in $k$-space to reduce the energy cost for the excitation around the $\pi$-point until condensation of these triplon excitations occur and order antiferromagnetically; (see Fig.~\ref{CondensationMechanismFigure}).
One recent work has tested this mechanism with dynamical mean field theory in the limit of infinite dimensions.\cite{Yunoki2016}
Here we ask the question: Can spin-F interactions from the $^4 S$ pathway cause condensation for large $J_H /U$? If so, then at which $k$-point does the condensation occur?
We show that for spin-F interactions there is a condensate but surprisingly the condensate does not always occur at the expected $k=0$ point. 

When $\lambda$ becomes much larger than the superexchange energy scale $t^2 / U$, the high energy $J_i = 2$ states become energetically unfavorable and can be ignored.
We project out $J_i = 2$ states from our spin-orbital superexchange Hamiltonians leaving just the $J_i = 0$ and $J_i = 1$ parts.
We utilize a set of operators $\cre{T}{i}$ to describe the triplon excitations from $J_i = 0$ states to $J_i = 1$ states.
These operators are defined by $\ket{J_i = 1, J_{i,z} = m} = \cre{T}{i,m} \ket{J_i = 0}$.
We then project the superexchange Hamiltonian onto the space of triplon operators, keeping only terms which are quadratic in the triplon operators (ie. $\cre{T}{i} \ann{T}{j}$, $\cre{T}{i} \cre{T}{j}$) and throw away terms with three and four triplon operators which constitute effective interactions between triplon excitations.
See Appendix \ref{CalcOfCondHam} for calculation details.
Since the projection of the magnetization operator is $\veck{M}_i = -i\sqrt{6} ( \ann{\veck{T}}{i} - \cre{\veck{T}}{i} ) - \tfrac{i}{2} \cre{\veck{T}}{i} \times \ann{\veck{T}}{i}$, the quadratic part describes interactions between van Vleck excitations.
This approximation cannot be justified deep in the condensed phase where interactions between excitations cannot be neglected, however, it can provide a good estimate of when the $t^2/U$ energy scale is large enough to support condensation and qualitatively what kind of magnetic ordering to expect.

There is, however, a more subtle consequence of the quadratic approximation.
Since each site may only accommodate at most one triplon, there is a hardcore Boson constraint on every site.
Although neglecting this constraint and the triplon-triplon interactions is necessary to put the solutions in closed form, the orbital \textit{frustration} from the $N_\mathrm{orb} = 2$ and $N_\mathrm{orb} = 1$ models is lost under these approximations.
This is separate from the orbital \textit{anisotropy} that will still remain present.
This is notable because anisotropic interactions usually cause frustration, yet here they will not due to the approximation that the triplons are non-interacting.
Again, deep in the condensed phase (the unexplored region of Fig.~\ref{AfmFmSchematicPhases}(b)), the exact solutions may qualitatively differ from the picture depicted here.

After projecting the superexchange Hamiltonian to this quadratic subspace and making a transformation to cubic coordinates (ie. $\cre{T}{-1} , \cre{T}{0} , \cre{T}{1} \rightarrow \cre{T}{x} , \cre{T}{y} , \cre{T}{z}$), the most general $\cre{T}{i} \ann{T}{j}$ term can be decomposed into the three parts given below
\begin{align}
H_{\mathrm{iso}}^{(ij)} & = J\, \veck{T}_i^\dagger \cdot \veck{T}_j + \mathrm{h.c.} \label{TriplonIsoForm} \\
H_{\mathrm{skew}}^{(ij)} & = \boldsymbol{D} \cdot \left( \veck{T}_i^\dagger \times \veck{T}_j \right)  + \mathrm{h.c.} \label{TriplonSkewForm} \\
H_{\mathrm{symm}}^{(ij)} & = \veck{T}_i^\dagger \cdot \boldsymbol{\Gamma} \cdot \veck{T}_j  + \mathrm{h.c.} \label{TriplonSymmForm}
\end{align}
and similarly for the $T_i^\dagger T_j^\dagger$ terms.
First consider the isotropic term in \eqref{TriplonIsoForm}.
When $J$ becomes large enough, all three flavors of triplons ($T_x^\dagger$, $T_y^\dagger$, $T_z^\dagger$) condense simultaneously.
Negative $J$ causes condensation at the $k$ point corresponding to a F condensate of van Vleck excitations while positive values cause condensation at the $\pi$ point corresponding to AF van Vleck excitations.
Like the Heisenberg term, it comes from the orbitally symmetric component of the interactions like those considered in \eqref{orbSymmChannelAll}.
The addition of further neighbor interactions can cause condensation at arbitrary $q$-vector.
Next, the skew symmetric term in \eqref{TriplonSkewForm} results in a magnetic spiral condensate at $q=\pm \pi/2$ points, and the addition of further neighbor interactions along with isotropic terms can make arbitrary $q$-spirals possible.
Like the Dzyaloshinskii-Moriya interaction, this term requires broken inversion symmetry.
Finally, the symmetric anisotropy in \eqref{TriplonSymmForm} picks one of the three flavors of triplons as the favored condensate due to orbital anisotropy.

This qualitative picture can be extended to finite temperature since the condensation mechanism falls into the Bose-Einstein condensation universality class.
Finite temperature condensation has been extensively studied,\cite{Batista2014} so we will instead focus the unique aspects of spin-orbital condensation from superexchange.

\subsection{Results}
\label{ResForNorb321}
We first determine which orbital geometries allow for F condensation when the hopping matrix is diagonal.
Then we decompose the hopping matrix into multipoles
\begin{equation}
t^{(ij)} = \sum_k t_0^k A_0^k
\label{condensateHoppingMatrix2}
\end{equation}
where multipoles are defined using Wigner-3j symbols
\begin{equation}
\bra{j m'} A_q^k \ket{jm} = (-1)^{j-m'} \left( \begin{array}{ccc} j & k & j \\ -m' & q & m \end{array} \right)
\end{equation}
and $t_0^k$ are the coefficients of the decomposition.
Then \eqref{Ht} is rewritten in the new form below.
\begin{equation}
H_{\mathrm{t}}^{(ij)} = 
\sum_{k} t_0^k \sum_{m \sigma} \left( A_0^k \right)_{mm} \, \cre{c}{im\sigma} \ann{c}{jm\sigma} 
\label{condensateHoppingOperator2}
\end{equation}
This form is particularly convenient to calculate the resulting spin-orbital superexchange form for each of the $t_0^k$ hopping matrices.
Here we will only give the results, and details of the calculation are relegated to Appendices~\ref{CalcOfCondHam} and \ref{condensationSuperexchangeSupp}.

For $N_\mathrm{orb} = 3$, the hopping matrix $t_{m'm}$ is simply proportional to $t_0^0$ in \eqref{condensateHoppingMatrix2} and \eqref{condensateHoppingOperator2}.
We find that isotropic hopping, $t_0^0$, only supports AF regardless of the value of Hund's coupling.
This result contradicts the claim of Ref.~\citenum{Meetei2015} that F condensation results for orbitally symmetric hopping.
If the hopping matrix is either skew-symmetric, $t_0^1$, or symmetrically anisotropic, $t_0^2$, the overall sign of the $^4 S$ pathways is the opposite to that of the $^2 D$ and $^2 P$ pathways, and large Hund's coupling can favor a F condensate.
The main difference between the isotropic term and the anisotropic terms is that both anisotropic terms feature matrix elements of different signs while the isotropic term does not.
Both the $N_\mathrm{orb} = 2$ and $N_\mathrm{orb} = 1$ models have hopping matrices described as linear combinations of $t_0^0$ and $t_0^2$.
However, only AF condensates result in these cases since their decompositions are closer to the isotropic hopping matrix instead of the anisotropic hopping matrix.
The ferromagetic condensation of triplon excitations is possible, but, due to the isotropic term only favoring AF, special orbital geometries are required for F condensation.

The lack of F condensation for most common orbital geometries has an immediate consequence on the phase diagram for $d^4$ materials.
In the limit of large spin-orbit coupling for \textit{any} value of $J_H / U$, there is a PM to AF transition with increasing superexchange.
However the limit of small $\lambda$ allows for both F and AF interactions.
Then there must be an additional phase transition between the AF condensate phase and a spin-orbital F phase at intermediate values of $(t^2/U) / \lambda$ when $J_H / U$ is large in the unexplored region of Fig.~\ref{AfmFmSchematicPhases}(b).

In the case of AF condensation, we give the critical condensation value $(t^2/U)/\lambda$ for each of the three models.
\begin{itemize}[leftmargin=*]
\item $N_\mathrm{orb} = 3$: Since this model possesses rotational invariance, each triplon flavor condenses simultaneously.
The effective singlet-triplet gap from on-site interactions is $\Delta = \lambda / 2 - \tfrac{z}{3} (t^2 / U)$ where $z$ is the coordination number and the inter-site interactions give $a_\delta = 5/3$ and $b_\delta = -4/3$ in \eqref{singleTriplonResult}.
Condensation occurs at the $(\pi,\pi,\pi)$ point at $t^2/U = \lambda / 40$.
\item $N_\mathrm{orb} = 2$: This model was studied in Ref.~\citenum{d4Khaliullin2013} where the orbital anisotropy was averaged away so that condensation occurred at $(\pi,\pi,\pi)$ at $t^2 / U = \lambda / 20$ for simple cubic lattices; without averaging 
it would occur at $(\pi,\pi,0)$ and equivalent directions.
We can conclude that triplon condensation is then likely to be active in both $4d$ and $5d$ transition metal oxides.
\item $N_\mathrm{orb} = 1$: In this case, condensation occurs along a degenerate set of points:  at $(k_x=2\pi/a, k_y=0, k_z)$  where $k_z$ can be arbitrary for the z-boson and the other 3 degenerate lines related by $C_4$ symmetry.
Here the 4 lines are parallel to the $k_z$ axis for the z boson, and the x and y bosons condense along lines being parallel to the $k_x$ and $k_y$ axes respectively. 
We find that $a_\delta = 1/6$ for directions perpendicular to a bond and $a_\delta = 2/3$ in the direction of a bond.
The values of $b_\delta$ are just the negatives of the $a_\delta$ values.
On a face-centered cubic lattice, we find a critical value of $t^2 / U= (3/32)\lambda$.
With this value, condensation is likely to occur in $4d$ compounds, but large values of spin-orbit coupling in $5d$ compounds will likely prevent condensation from occurring.

\end{itemize}

\subsection{Local Interactions versus Condensation}
It is surprising that although local F interactions were found for large values of $J_H / U$, F condensation did not appear even in the isotropic $N_\mathrm{orb} = 3$ model which was free of orbital frustration.
Even more surprising was that isotropic hopping is the cause of this unexpected result despite our calculations in Figs.~\ref{conservedJphasediagram}(b).
This discrepancy can be resolved by examining the two site problem more carefully, and our goal is to tie the two site and lattice condensation results together.
To do this, we focus on the key problem presented: the lack of F condensation in a spin-orbital superexchange Hamiltonian which is explicitly F by construction.

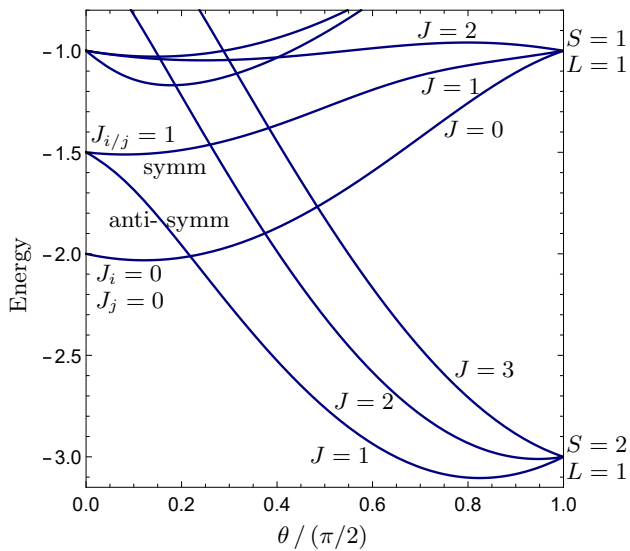
\begin{figure}
\vspace{0.1cm}

\begin{overpic}[width=2.8in,clip]{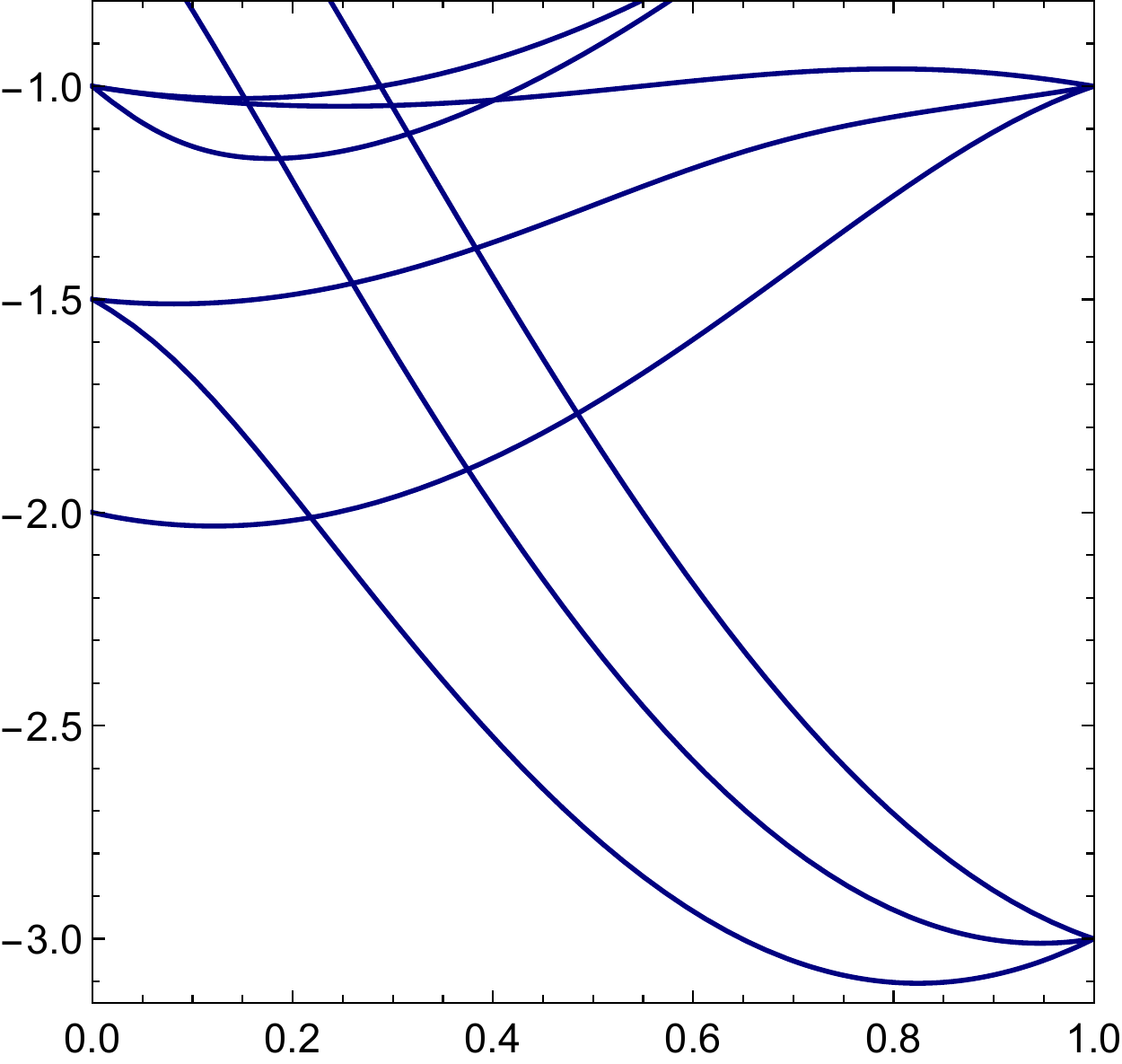}
 \put(44,-5){$\theta \, / \, (\pi / 2)$}
 \put(-6,38){\rotatebox{90}{Energy}}
 
 \put(98,12){$S=2$}
 \put(98,7){$L=1$}
 
 \put(98,88){$S=1$}
 \put(98,83){$L=1$}
 
 \put(77,26){$J=3$}
 \put(54.5,20){$J=2$}
 \put(50,10){$J=1$}
 
 \put(70,89.5){$J=2$}
 \put(70.7,79){$J=1$}
 \put(75,71){$J=0$}
 
 \put(10,44){$J_i = 0$}
 \put(10,39){$J_j = 0$}

 \put(9,70){$J_{i/j} = 1$}
 
 \put(19,64){symm}
 \put(12.5,54){anti- symm}
\end{overpic}

\vspace{0.3cm}

\caption{
Energy levels of the effective Hamiltonian $H_\mathrm{SE} + H_\mathrm{SOC}$ appearing in \eqref{HSElastsection} and \eqref{HSOClastsection} with the parameterization $\lambda = \cos \theta$ and $J_\mathrm{SE} = \sin \theta$. The levels are labeled by their good quantum numbers. In the $\theta = 0$ limit, the eigenstates of spin-orbit coupling are used, and, in the $\theta = \pi/2$ limit, the eigenstates of the spin-orbital superexchange Hamiltonian are used. The interpretations of the states are discussed in the main text.
\label{energyLevelStructureCondensation}
}
\end{figure}

First we will rewrite the ferromagnetic part of the superexchange Hamiltonian for the $N_\mathrm{orb} = 3$ model appearing in \eqref{effectiveSEHamiltonian} and \eqref{orbSymmChannelA}
\begin{equation}
H_{\mathrm{SE}} =  - \frac{J_\mathrm{SE}}{2} \left(2 + \veck{S}_i \cdot \veck{S}_j \right) \left[ 2 - \veck{L}_i \cdot \veck{L}_j -  (\veck{L}_i \cdot \veck{L}_j)^2 \right]
\label{HSElastsection}
\end{equation}
which is pitted against the lowest order spin-orbit correction.
\begin{equation}
H_\mathrm{SOC} = \frac{\lambda}{2} \left( \veck{L}_i \cdot \veck{S}_i + \veck{L}_j \cdot \veck{S}_j \right)
\label{HSOClastsection}
\end{equation}
Fig.~\ref{energyLevelStructureCondensation} shows the energy spectra for a two site system parameterized with $\lambda = \cos \theta$ and $J_\mathrm{SE} = \sin \theta$.
On the right hand side at $\theta = \pi / 2$, the lowest energy levels are the total $S=2$ and $L=1$ states.
A small amount of spin-orbit coupling splits the states into total $J=1,2,3$ states as already stated in Section \ref{3activeOrbitalSection}.
On the left hand side at $\theta = 0$, the two lowest energy levels are as follows: a non-degenerate state with both sites in the non-magnetic $J=0$ singlet state and a six-fold degenerate first excited level where one of the two sites contains an excitation.
These levels correspond to the vacuum and the triplon band of excitations in the condensation picture.
Introducing a small amount of superexchange splits the first excited states into symmetric and anti-symmetric states.
The lower energy states are the anti-symmetric ones which correspond to condensation at the $\pi$-point from a triplon condensation Hamiltonian $\cre{\veck{T}}{i} \cdot \ann{\veck{T}}{j} + \mathrm{h.c.}$ with a positive hopping coefficient.

Here lies the source of the discrepancy between two site results and the lattice condensation result.
There are two sides to the lowest energy $J=1$ ($S=2$, $L=1$) level: the regime where $\lambda$ and $J_\mathrm{SE}$ are comparable and the regime where $\lambda \gg J_\mathrm{SE}$ ($\theta \approx 0$).
Section \ref{3activeOrbitalSection} showed that the regime where the two interactions were comparable produced F spin interactions.
However these $J=1$ states in the $\lambda \gg J_\mathrm{SE}$ regime correspond to triplon AF.
Even though the two regimes are smoothly connected, the nonmagnetic $J_i = 0$ states are lower in energy than the triplon excitations.
Then two site exact diagonalization covers up this aspect of AF while the lattice limit allows the anti-symmetric level (triplon band) decrease enough in energy to reveal the AF nature of the $\theta \approx 0$ part of the lowest energy $J=1$ states.

To summarize, while from the exact diagonalization, it seemed like the $J=1$ line in Fig.~\ref{energyLevelStructureCondensation} should have been FM for all $\theta$.  However, 
in the region where it was not FM and was actually the anti-symmetric AFM triplon condensate (at the two-site level), a different AFM energy level $(J_i=J_j=0)$ was the one exact diagonalization was measuring, not the $J=1$ AFM condensate line which was the next-lowest energy level.  
In the full lattice (not 2 site), the anti-symmetric level is the condensate (AFM) and drops below the $(J_i=J_j=0)$ energy level.  This arises because on the right side for $\theta$ greater than the crossing, $J=1$ comes from $S=2$ and $L=2$ (FM) while on the left side $J=1$ comes from an antisymmetric splitting of $J_i=1$ and $J_j=0$  with  $J_i=0$ and $J_j=1$ (AFM).

Lastly Fig.~\ref{energyLevelStructureCondensation} highlights a difference between local spin AF and triplon AF.
The exact diagonalization results in Figs.~\ref{conservedJphasediagram} show that the two site model features a smooth transition between $J_i = 0$ states and AF behavior.
The same behavior is found in Fig.~\ref{energyLevelStructureCondensation} by following the $J_i = 0$ line to the $J=0$ line which is the transition from local non-magnetic singlets to a total $J=0$ singlet between sites.
(For reference, the AF-F level crossing discussed in the first two sections would be shown here by this AF $J=0$ ($S=1$, $L=1$) level being overtaken by the F $J=1$ ($S=2$, $L=1$) level when the other two pathways in \eqref{effectiveSEHamiltonian} are included.)
The AF from triplon condensation corresponds to the anti-symmetric level, and these two types of AF are therefore different.
In fact, from the quantum numbers shown in Fig.~\ref{energyLevelStructureCondensation}, the two types of AF belong to different irreducible representations at the two site level.
It is then possible that there is an additional phase boundary in Fig.~\ref{AfmFmSchematicPhases}(b) which separates AF triplon BEC and spin AF despite the fact that they are both AF phases.

\section{Materials and Experiments}
\label{MaterialsExperimentsSection}
The $d^5$ irridates (Ir$^{4+}$) with half filled $j=1/2$ bands have attracted a significant amount of attention due to the interplay between strong correlations and spin-orbit coupling.
Although experimental studies have recently been focused on these materials, many strongly correlated oxides with moderate to strong spin-orbit coupling have $d^4$ configurations.
A well studied example\cite{Nakatsuji1997,Braden1998} is Ca$_2$RuO$_4$ which necessarily violates Hund's rules which require non-magnetic $J_i=0$ Ru sites, however there are many less well studied $d^4$ materials.
There are many double perovskites of the form A$_2$BB'O$_6$ where both A and B have completely filled valence shells and the B' site is in the $d^4$ configuration including La$_2$ZnRuO$_6$,\cite{BattleFrostKim1995} La$_2$MgRuO$_6$,\cite{DassYanGoodenough2004} Sr$_2$YIrO$_6$,\cite{CaoKaul2013} Ba$_2$YIrO$_6$,\cite{Kennedy2015,PhelanCava2016,Buchner2016} and a large array of compounds with the form Sr$_2B$IrO$_6$.\cite{MarcoHaskel2015}
However, from our $N_\mathrm{orb} = 1$ results in the previous sections, $5d$ double perovskites are unlikely candidates for triplon condensation due to the small superexchange energy scales when compared to spin-orbit coupling.
Additionally, it has been suggested that the observed magnetism in $5d$ double perovskites is due to disorder and/or impurities.\cite{Buchner2016}
However, both $4d$ compounds and compounds with more than one active orbital ($N_\mathrm{orb} = 2$ or $3$) should be good candidates.
Honeycomb $d^4$ oxides Li$_2$RuO$_3$ and Na$_2$RuO$_3$ have been found to order antiferromagnetically.\cite{WangCao2014}
Other $d^4$ oxides include post-perovskite\cite{Bremholm2011} NaIrO$_3$ and pyrochlore\cite{ZhaoYan2016} Y$_2$Os$_2$O$_7$.

Many probes can be used to deduce the existence of novel magnetism in $d^4$ materials.
In particular, both magnetic susceptibility and x-ray absorption spectroscopy (XAS) have the advantage that magnetic ordering is not required to infer the existence of moments.
The first test for novel magnetism in $d^4$ systems comes at the level of magnetic susceptibility measurements.
Curie Weiss fits provide a measure of the effective magnetic moment, and measuring a non-zero effective moment (of order 1 $\mu_B$) is a direct indication that the ground state is not the product of non-magnetic $J_i=0$ singlets.
Determining whether the ground state is not the product of non-magnetic singlets can also be probed by XAS.\cite{MarcoVeenendaal2010}
In the non-magnetic $J_i=0$ singlet state, $L_2$ absorption edge intensity is zero while the $L_3$ edge is non-zero which leads to a diverging branching ratio
\begin{equation}
\mathrm{B.R.} = \frac{I_{L_3}}{I_{L_2}} = \frac{2+r}{1-r}
\end{equation}
where $r = \langle (-\veck{L}_i) \cdot \veck{S}_i \rangle / \langle n_h \rangle$.
However, the $J_i=1$ and $J_i=2$ states lead to finite $L_2$ edges with magnitudes comparable of that of the $L_3$ edge.
Thus measuring a branching ratio of order 1 is direct evidence against the non-magnetic singlet ground state.

\section{Conclusions}
We have studied how superexchange opposes the effect of spin-orbit coupling in $d^4$ systems and induces local moments and interactions between them.
If Hund's coupling is large, the local interactions favor ferromagnetism instead of the expected antiferromagnetism.
We also found that at least two orbitals need to be involved for this local ferromagnetic behavior to be energetically favorable.
The condensation mechanism allows AF to generally be favorable in both $4d^4$ and $5d^4$ compounds.
However because isotropic orbital interactions favor antiferromagnetic condensation regardless of how large Hund's coupling is, ferromagnetic condensation is unlikely in materials systems.
We would like to highlight again that the ability of rotationally invariant atomic spin-orbit coupling to flip the sign of the effective exchange constant on a superexchange interaction is a unique feature of spin-orbital systems and has no analog in pure spin systems.

The effective magnetic Hamiltonians derived here for transition metal oxides with $d^4$ occupancy can be directly used for the particle-hole symmetric $d^2$ occupancy as well after changing the sign of the hopping and spin-orbit couplings. These Hamiltonians lay the foundation for spin-orbit coupled Hamiltonians in the $t_{2g}$ sector. Going forward, different analytical and numerical methods can now be applied to obtain detailed phase diagrams. 

\section{Acknowledgments}
We thank Jiaqiang Yan for useful discussions.
We acknowledge the support of the CEM, an NSF MRSEC, under grant DMR-1420451.

\bibliography{include}

\appendix
\section{Coulomb Interaction}
The Coulomb interaction for $t_{2g}$ orbitals is derived here.\cite{Kanamori1963,doi:10.1146/annurev-conmatphys-020911-125045}
The three $t_{2g}$ orbitals are labeled by  $\lbrace \phi_a, \phi_b, \phi_c \rbrace$ where $(a,b,c)$ is a permutation of $(yz,zx,xy)$.
While these orbitals are not required to be atomic orbitals, we still require them to be invariant under the $O_h$ group.

We now give the full on-site Coulomb interaction in terms of orbital indices $\lbrace \alpha,\beta,k,\zeta \rbrace$ and spin indices $\lbrace \sigma_1,\sigma_2,\sigma_3,\sigma_4 \rbrace$.
\begin{equation}
H_{\mathrm{int}} = \sum_{\alpha \sigma_1} \sum_{\beta \sigma_2} \sum_{k \sigma_3} \sum_{\zeta \sigma_4} \pcre{\alpha \sigma_1} \pcre{\beta \sigma_2} \;  V_{\alpha \sigma_1 , \beta \sigma_2}^{k \sigma_3 , \zeta \sigma_4}  \; \pann{\zeta \sigma_4} \pann{k \sigma_3} 
\end{equation}
\begin{equation}
V_{\alpha \sigma_1, \beta \sigma_2}^{k \sigma_3, \zeta \sigma_4} = \bra{\alpha \beta} V \ket{k \zeta} \delta_{\sigma_1}^{\sigma_3} \delta_{\sigma_2}^{\sigma_4} - \bra{\alpha \beta} V \ket{\zeta k} \delta_{\sigma_1}^{\sigma_4} \delta_{\sigma_2}^{\sigma_3}
\end{equation}
\begin{multline}
\bra{\alpha \beta} V \ket{k \zeta} = \int \dif \veck{r}_1 \, \dif \veck{r}_2 \; \phi_\alpha (\veck{r}_1)  \phi_\beta (\veck{r}_2) V(\veck{r}_1 - \veck{r}_2) \\ \times \phi_k (\veck{r}_1) \phi_\zeta (\veck{r}_2)
\end{multline}
We require the effective on-site Coulomb interaction to be invariant under all three mirror plane operations.
The four non-zero quantities characterizing $V$ in the $t_{2g}$ subspace are then given below.
We use $J_H$ instead of $J$ to avoid confusion with total angular momentum.
\begin{equation}
\begin{split}
\bra{aa} V \ket{aa} \equiv U & \hspace{1cm} \bra{ab} V \ket{ab} \equiv U' \\
\bra{ab} V \ket{ba} \equiv J_H & \hspace{1cm} \bra{aa} V \ket{bb} \equiv J_H' 
\end{split}
\end{equation}
We will expand $H_{\mathrm{int}}$ using these four parameters.
It will be convenient to group group terms based on the five types of $V_{\alpha \sigma_1 , \beta \sigma_2}^{k \sigma_3 , \zeta \sigma_4}$ terms encountered.
\begin{equation}
\begin{array}{lcl}
V_{a \uparrow, a \downarrow}^{a \uparrow , a \downarrow} = - V_{a \uparrow, a \downarrow}^{a \downarrow , a \uparrow}  = U 
\vspace{0.15cm}
&
\hphantom{blah}
&
V_{a \uparrow, b \uparrow}^{a \uparrow, b \uparrow} = U' - J_H
\\
V_{a \uparrow, a \downarrow}^{b \uparrow, b \downarrow } = - V_{a \uparrow, a \downarrow}^{b \downarrow, b \uparrow }  = J_H'
\vspace{0.15cm}
&
&
V_{a \uparrow, b \downarrow}^{a \uparrow, b \downarrow } = U' 
 \\
&
&
V_{a \uparrow, b \downarrow}^{a \downarrow , b\uparrow} = -J_H
\end{array}
\end{equation}
The final result is given below
\begin{multline}
H_{\mathrm{int}} =  U N_{\uparrow \downarrow} + U' N_{\uparrow \downarrow}'  + (U'-J_H) N_{\uparrow \uparrow}'  
\\
 -J_H \, H_{\mathrm{ex}} + J_H' \, H_{\mathrm{pair}}
\label{Antoine3}
\end{multline}
with:
\begin{equation}
N_{\uparrow \downarrow} = \sum_{a} n_{a\uparrow} n_{a\downarrow}
\end{equation}
\begin{equation}
N_{\uparrow \downarrow}' = \sum_{a \neq b}  n_{a\uparrow} n_{b \downarrow}
\end{equation}
\begin{equation}
N_{\uparrow \uparrow}' = \tfrac{1}{2}\sum_{a \neq b} \sum_{\sigma} n_{a \sigma} n_{b \sigma}
\end{equation}
\begin{equation}
H_{\mathrm{ex}} =  \sum_{a \neq b} \pcre{a\uparrow} \pcre{b\downarrow} \pann{b \uparrow} \pann{a \downarrow}
\end{equation}
\begin{equation}
H_{\mathrm{pair}} = \sum_{a\neq b} \pcre{a\uparrow} \pcre{a\downarrow} \pann{b \downarrow} \pann{b \uparrow} 
\end{equation}
We will eliminate $N_{\uparrow \downarrow}$, $N_{\uparrow \downarrow}'$, and $N_{\uparrow \uparrow}'$ in favor of the number, spin, and angular momentum operators below.
\begin{equation}
N = \sum_{a \, \sigma} n_{a \sigma}
\end{equation}
\begin{equation}
\veck{S} = \tfrac{1}{2} \sum_{a} \sum_{\sigma \, \sigma '}  \pcre{a \sigma} \veck{\tau}_{\sigma \sigma '} \pann{a \sigma ' }
\end{equation}
\begin{equation}
L_a = i \sum_{b \, c} \sum_{\sigma} \epsilon_{abc} \pcre{b \sigma} \pann{c \sigma}
\end{equation}
Using the preceding definitions, we obtain the following relations
\begin{align}
\tfrac{1}{2} N (N-1) & = N_{\uparrow \downarrow} + N_{\uparrow \uparrow}' + N_{\uparrow \downarrow}' \\
\tfrac{1}{4} N (N-1) + \tfrac{3}{4} N + H_{\mathrm{ex}} -S^2 & = 2 N_{\uparrow \downarrow} + N_{\uparrow \downarrow}' \\
N - H_{\mathrm{ex}} - H_{\mathrm{pair}} -\tfrac{1}{2} L^2 & = N_{\uparrow \uparrow}'
\end{align}
which can be used to rewrite the Hamiltonian in the following form.
\begin{multline}
H_{\mathrm{int}}  = \tfrac{1}{4} (3 U' - U) N(N-1) + (\tfrac{7}{4} U - \tfrac{7}{4} U' - J_H) N  \\
 + (U' - U) S^2 + \tfrac{1}{2}(U' - U + J_H) L^2 \\
 + (U' - U + J_H + J_H') P 
\label{Antoine5}
\end{multline}
Requiring $J_H = J_H'$, the relation $U = U' + 2J_H$ gives the rotationally invariant form of $H_{\mathrm{int}}$ used in \eqref{Ht2g}.

\section{Effective Hamiltonian}
\label{effectiveHamiltonianAppendix}
We calculate the effective spin-orbit coupling in the $^3 P$ subspace.
The linear correction for spin-orbit coupling is given by 
\begin{equation}
H_{\mathrm{SOC},(1)} = \frac{ \bra{ ^3 P |} H_{\mathrm{so}} \ket{| ^3 P} }{ \bra{ ^3 P |} \veck{L} \cdot \veck{S} \ket{| ^3 P} } \veck{L} \cdot \veck{S}
\end{equation}
where $\bra{ ^3 P |} H_{\mathrm{so}} \ket{| ^3 P}$ and $\bra{ ^3 P |} \veck{L} \cdot \veck{S} \ket{| ^3 P}$ are the reduced matrix elements of the two operators respectively.
Evaluating this ratio, we obtain $\bra{ ^3 P |} H_{\mathrm{so}} \ket{| ^3 P} / \bra{ ^3 P |} \veck{L} \cdot \veck{S} \ket{| ^3 P} = \lambda / 2$.

Now we calculate second order energy corrections for the $^3 P$ levels due to spin-orbit coupling.
Since $H_{\mathrm{so}}$ conserves total angular momentum, only energy levels of the same total angular momentum $J$ are coupled together.
Then $^3 P_2$ couples to $^1 D_2$ and $^3 P_0$ couples to $^1 S_0$.
The $^3 P_1$ level remains unshifted.

The second order correction for $^3 P_2$ requires us to calculate the matrix elements $\bra{^1 D_2} H_{\mathrm{so}} \ket{^3 P_2}$ for the $^3 P_2$ energy shift, and it suffices to calculate the $\bra{^1 D_2, J_z = +2} H_\mathrm{so} \ket{^3 P_2, J_z = +2}$ matrix element given below
\begin{multline}
\bra{\mathrm{vac}} \ann{c}{1\uparrow} \ann{c}{1\downarrow}
\left(
\lambda \cre{c}{m\sigma} 
\left( \veck{l}_{mm'} \cdot \veck{s}_{\sigma \sigma'} \right)
\ann{c}{m'\sigma'}
\right)
\cre{c}{0\uparrow} \cre{c}{1\uparrow} \ket{\mathrm{vac}}
\end{multline}
which is only nonzero for $\veck{l}_{10} \cdot \veck{s}_{\downarrow \uparrow} = \tfrac{1}{2} l_{10}^+  s_{\downarrow \uparrow}^- = 1/\sqrt{2}$.
Then with an energy denominator of $E (^1 D) - E(^3 P) = 2J_H$ , we have a second order energy shift of:
\begin{equation}
\Delta E^{(2)} \left(^3 P_2 \right) = - \frac{ \left( \lambda / \sqrt{2} \right)^2 }{ 2 J_H }
\end{equation}
Similarly we can calculate the matrix element $\bra{^1 S_0} H_{\mathrm{so}} \ket{^3 P_0} = \lambda \sqrt{2}$, and, with an energy denominator of $E (^1 S) - E(^3 P) = 5 J_H$, we obtain:
\begin{equation}
\Delta E^{(2)} \left( ^3 P_0 \right) = - \frac{ \left( \lambda \sqrt{2} \right)^2 } { 5 J_H }
\end{equation}
This is represented in operator form below.
\begin{equation}
H_{\mathrm{SOC},(2)} = \frac{ \lambda^2 }{J_H} \left( \frac{1}{20} -\frac{1}{8} \veck{L}\cdot \veck{S} - \frac{7}{40} \left( \veck{L} \cdot \veck{S} \right)^2 \right)
\end{equation}
The total effective spin-orbit interaction used in \eqref{effectiveSOCHamiltonian} is just $H_{\mathrm{SOC},(1)} + H_{\mathrm{SOC},(2)}$.

The effective superexchange Hamiltonian is broken into three parts based on the energy value of the intermediate $d^3$ configuration in the process $d^4 d^4 \rightarrow d^3 d^5 \rightarrow d^4 d^4$.
Using \eqref{Ht2g}, energies of the relevant states for computing this Hamiltonian are given below.
\begin{equation}
E[d^4(^3 P),d^4(^3 P)] = 12 U - 26 J_H
\end{equation}
\begin{equation}
\begin{split}
E[d^3(^4 S),d^5] & = 13 U - 29 J_H \\
E[d^3(^2 D),d^5] & = 13 U - 26 J_H \\
E[d^3(^2 P),d^5] & = 13 U - 24 J_H 
\end{split}
\end{equation}
The three energy differences, $E(d^3,d^5) - E(d^4,d^4)$, are given below.
\begin{equation}
\begin{split}
E[d^3(^4 S),d^5] - E[d^4(^3 P),d^4(^3 P)] & = U - 3J_H \\
E[d^3(^2 D),d^5] - E[d^4(^3 P),d^4(^3 P)] & = U \\
E[d^3(^2 P),d^5] - E[d^4(^3 P),d^4(^3 P)] & = U + 2J_H 
\end{split}
\label{appendixEnergyDifferences}
\end{equation}
Using these energy differences, $H_{\mathrm{SE}}$ is computed from \eqref{Ht2g} and \eqref{Ht} using the following scheme.
\begin{equation}
H_{\mathrm{SE}} = - \mathbb{P} H_{\mathrm{t}} \frac{1}{H_{\mathrm{int}} - E[d^4(^3 P),d^4(^3 P)]} H_{\mathrm{t}} \mathbb{P}
\label{howToPertTheory}
\end{equation}
Since the denominator must take on one of the three energy differences in \eqref{appendixEnergyDifferences}, we can separate \eqref{howToPertTheory} into the three pathways.
\begin{multline}
H_{\mathrm{SE}} = - \left( \tfrac{t^2}{U - 3J_H} \right) H_{\mathrm{SE}, ( ^4 S)} - \left( \tfrac{t^2}{U} \right) H_{\mathrm{SE}, (^2 D)} 
\\
- \left( \tfrac{t^2}{U + 2J_H} \right) H_{\mathrm{SE},( ^2 P )} 
\end{multline}
Spin symmetric interactions constrain each pathway to the form
\begin{equation}
H_{\mathrm{SE},(\xi)}^{(ij)} = \left( \alpha_{\xi} + \beta_{\xi} \veck{S}_i \cdot \veck{S}_j \right) \mathcal{O}_{ij}^{\xi}
\end{equation}
where $\alpha$ and $\beta$ are real coefficients and $\mathcal{O}_{ij}$ is an orbital interaction determined by $t_{m'm}^{(ij)}$.

\section{Condensation Formalism}
\label{CalcOfCondHam}
Here we derive the general triplon condensation Hamiltonian from a spin-conserving superexchange Hamiltonian.
Since each superexchange pathway can be written as the product of orbital interactions and spin interactions as in \eqref{effectiveSEHamiltonian}, the effective pathways can be decomposed into the product of orbital and spin multipole operators on each site.
\begin{equation}
\begin{split}
H^{(ij)} = \sum_{l l' m m'} x_{m m'}^{l l'} ({L}_i)_{m}^{l} ({L}_j)_{m'}^{l'} \sum_{s s' \sigma \sigma' } y_{\sigma \sigma'}^{s s'} ({S}_i)_{\sigma}^{s} ({S}_j)_{\sigma'}^{s'}
\end{split}
\end{equation}
Here ${L}_i$ and ${S}_i$ are the multipole operators for the orbital and spin parts of site $i$, and $x_{mm'}^{ll'}$ and $y_{\sigma \sigma '}^{s s'}$ are coefficients of the decomposition with $l \in \lbrace 0, 1, 2\rbrace$ and $-l \le m \le l$, etc.
We rewrite the superexchange Hamiltonian using total orbital operators $\mathcal{O}$ and total spin operators $\mathcal{S}$.
\begin{equation}
H^{(ij)} = \sum_{ll' LM} \alpha_M^{ll'L} {\mathcal{O}}_M^{ll'L}
 \sum_{ss' S\Sigma} \beta_{\Sigma}^{ss' S} \mathcal{S}_{\Sigma}^{ss'S}
\label{generalSuperexchangeForm}
\end{equation}
\begin{equation}
\mathcal{O}_M^{ll'L} =  \sum_{mm'} \left\langle lm,l'm' | LM \right\rangle  ({L}_i)_m^l ({L}_j)_{m'}^{l'}
\label{orbitalAngularMomentumFromLiLj}
\end{equation}
\begin{equation}
\mathcal{S}_\Sigma^{ss'S} =  \sum_{\sigma \sigma'} \left\langle s\sigma,s'\sigma' | S\Sigma \right\rangle  ({S}_i)_\sigma^s ({S}_j)_{\sigma'}^{s'}
\end{equation}
The symbol $\langle lm,l'm' | LM \rangle$ is a Clebsch-Gordon coefficient, and $\alpha_M^{ll' L}$ and $\beta_\Sigma^{ss' S}$ are the new coefficients of the decomposition.
Since $S = \Sigma = 0$ for superexchange interactions preserving spin symmetry as in \eqref{effectiveSEHamiltonian}, we have the following superexchange interaction.
\begin{equation}
H^{(ij)} = \sum_{ll'LM} \alpha_M^{ll'L} \mathcal{O}_M^{ll'L}
 \sum_{s} \beta_{0}^{ss0} \mathcal{S}_{0}^{ss 0} 
\label{generalMultipoleForm}
\end{equation}

\begin{figure}
\begin{overpic}[width=8.7cm]{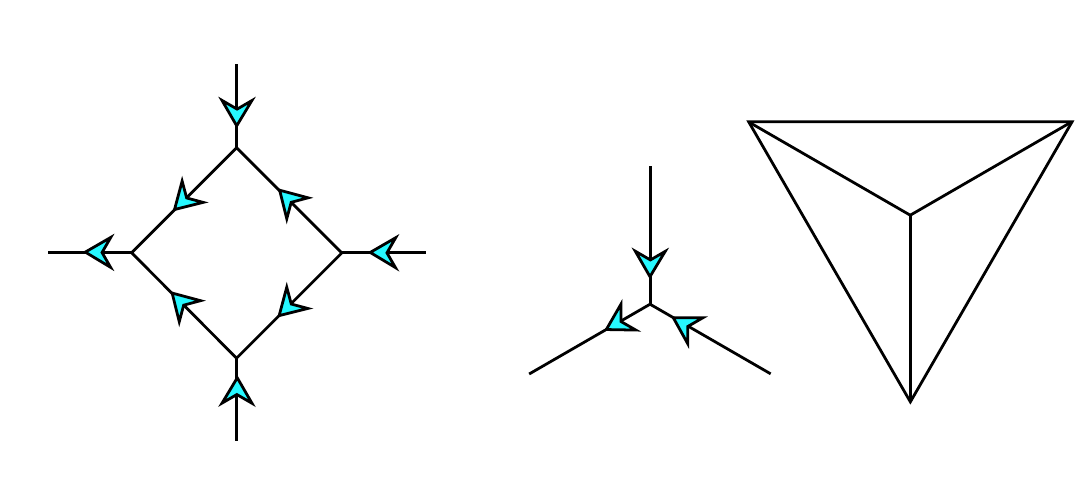}
 \put(-0.5,41){\textbf{(a)}}
 \put(0,21.7){$1\zeta$}
 \put(40,21.7){$0\hspace{0.4cm}=$}
 \put(20,3){$s\sigma$}
 \put(20,41){$lm$}
 \put(14,14){1}
 \put(14,29){1}
 \put(28,14){1}
 \put(28,29){1}
 \put(45.5,9){$1\zeta$}
 \put(71,9){$s\sigma$}
 \put(58,32){$lm$}
 \put(73,21){$1$}
 \put(93,21){$1$}
 \put(83,36){$1$}
 \put(81,21){$1$}
 \put(79,30){$l$}
 \put(87,30){$s$}
\end{overpic}
\begin{overpic}[width=8.7cm]{angular_momentum_figure.pdf}
 \put(-0.5,41){\textbf{(b)}}
 \put(-0.8,21.7){$1m'$}
 \put(40,21.7){$1m\hspace{0.25cm}=$}
 \put(20.8,2.4){$0$}
 \put(19.0,41){$LM$}
 \put(14,14){$s$}
 \put(14,29){$l'$}
 \put(28,14){$s$}
 \put(28,29){$l$}
 \put(45,9){$1m'$}
 \put(71,9){$1m$}
 \put(57,32){$LM$}
 \put(73,21){$l'$}
 \put(93,21){$l$}
 \put(83,36){$s$}
 \put(81,21){$L$}
 \put(79,30){$1$}
 \put(87,30){$1$}
\end{overpic}
\caption{
\coloronline The flow of angular momentum is graphically shown where ingoing arrows are incoming angular momentum and outgoing arrows are outgoing angular momenta.
Wigner-3j symbols and Clebsch-Gordan coefficients are vertices with three legs while the scalar contraction of four Wigner-3j symbols (right) is a Wigner-6j symbol.\cite{YLV1962,Brink1962angular}
(a) Equation \eqref{LSprojectToTDaggerT} is shown in graphical form for the $\cre{T}{\zeta}$ part of the equation.
A $J=0$ state is decomposed into its $L=1$ and $S=1$ components which are acted on by the $(L_i)_m^l$ and $(S_i)_\sigma^s$ operators.
The resulting $L=1$ and $S=1$ are combined together to give a $J=1$ state with quantum number $\zeta=m+\sigma$.
(b) The projection of equation \eqref{generalMultipoleForm} to \eqref{generalMultipoleHopping} conserves angular momentum.
Equation \eqref{generalMultipolePairing} will appear similarly except that $1m'$ and $1m$ add to yield $LM$ instead.
\label{AngularMomentumFigure}
}
\end{figure}
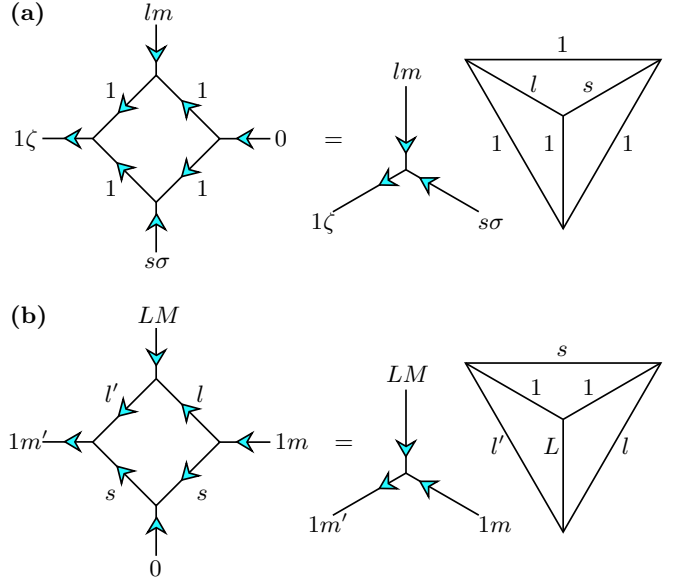

We now project out the high energy $J_i = 2$ states from the Hamiltonian and only leave the $J_i = 0$ and $1$ components.
Since we are only interested in the quadratic part of the result which couples sites together (ie. $\cre{T}{i} \ann{T}{j}$, $\cre{T}{i} \cre{T}{j}$) and captures the condensation of triplon excitations, we project directly to this subspace and ignore terms like $\cre{T}{i} \ann{T}{i}$ since they only amount to energy shifts.
This projection is accomplished by first projecting the spin-orbital operators on each site to the space of triplon creation and annihilation operators.
The Wigner-Eckart theorem requires this projection of the product of multipole operators $(L_i)_m^l (S_i)_\sigma^s$ is proportional to $\cre{T}{i,\zeta}$ and $\ann{T}{i,-\zeta}$ where $\zeta = m+\sigma$,
\begin{multline}
(L_i)_m^l (S_i)_{\sigma}^{s} 
\rightarrow
\left(
\begin{array}{ccc}
1 & l & s \\
-\zeta & m & \sigma
\end{array}
\right)
\left\lbrace 
\begin{array}{ccc}
1 & l & s \\
1 & 1 & 1
\end{array}
\right\rbrace
\\
\times
\left[
(-1)^{l+1-\zeta} \,
\cre{T}{i,\zeta}
+
(-1)^{s} \,
\ann{T}{i,-\zeta}
\,
\right] 
\label{LSprojectToTDaggerT}
\end{multline}
and the factor in braces is a Wigner-6j symbol.
The flow of angular momentum due to the $\cre{T}{\zeta}$ operator from this projection is represented graphically in Fig.~\ref{AngularMomentumFigure}(a).

Now we combine the terms from sites $i$ and $j$ with the constraint from \eqref{generalMultipoleForm}.
It will be more convenient to project each type of term appearing in \eqref{generalMultipoleForm} individually, so we define the operator $H' = \mathcal{O}_{M}^{ll'L} \mathcal{S}_0^{ss0}$.
Note that $H'$ is not generally Hermitian, but, for every $LM$ term, there is a complementing $L(-M)$ term so that $H^{(ij)}$ in \eqref{generalMultipoleForm} is Hermitian.
Then the projection of $H'$ takes the two forms below.
\begin{equation}
H_{\cre{T}{i} \ann{T}{j}}'
=
 g_1
\sum_{m'm} \cre{T}{i,m'} \sqrt{2L+1} \bra{1m'} A_M^L \ket{1m} \ann{T}{j,m}
\label{generalMultipoleHopping}
\end{equation}
\begin{equation}
H_{\cre{T}{i} \cre{T}{j}}'
=
 g_2
\sum_{m'm}
\cre{T}{i,m'} \left\langle 1m',1m | LM \right\rangle \cre{T}{j,m}
\label{generalMultipolePairing}
\end{equation}
The $\bra{1m'} A_M^L \ket{1m}$ coefficients are the multipole matrix elements previously defined.
The coefficients $g_1 = (-1)^{l'} g$ and $g_2 = (-1)^{-s} g$ are given below.
\begin{equation}
g =
\frac{ (-1)^{l + L + 1} }{ \sqrt{2s+1} } 
\left\lbrace \begin{array}{ccc}
1 & L & 1 \\ l & s & l'
\end{array} \right\rbrace
\left\lbrace \begin{array}{ccc}
1 & l & s \\ 1 & 1 & 1
\end{array} \right\rbrace
\left\lbrace \begin{array}{ccc}
1 & l' & s \\ 1 & 1 & 1
\end{array} \right\rbrace
\end{equation}

It is important to note that the angular momentum contained in $H'$ (and also $H^{(ij)}$) is conserved under projection.
Furthermore, after making the transformation to cubic coordinates, we immediately have our result: the $L=0$ component gives the coefficient $J$ in \eqref{TriplonIsoForm}, the $L=1$ components give $\veck{D}$ in \eqref{TriplonSkewForm}, and the $L=2$ components give the irreducible components of $\boldsymbol{k}$ in \eqref{TriplonSymmForm}.
This is the advantage of writing the orbital and spin parts using their total angular momenta.

While we have restricted the calculation to spin-symmetric superexchange interactions, the concept applies more generally.
If the total spin operators are non-trivial, further combine the total orbital and total spin operators into a total spin-orbital operator.
The total angular momentum contained in the total spin-orbital operator will be that which appears in the condensation Hamiltonian.

Now we determine the critical value of $(t^2/U) / \lambda$ where condensation occurs.
To simplify, we consider the condensation of a single flavor of triplon.
Then the condensation Hamiltonian has the following form.
\begin{equation}
H = \Delta \sum_i \cre{T}{i}\ann{T}{i} + \frac{t^2}{U} \sum_{i,\delta}
\left( \frac{a_\delta}{2} \cre{T}{i+\delta}\ \ann{T}{i} +  \frac{b_\delta}{2} \ann{T}{i + \delta}\ann{T}{i} + \mathrm{h.c.} \right)
\label{singleTriplonCondensation}
\end{equation}
Here we have assumed that the singlet-triplet gap is given by $\Delta = \lambda / 2 + \nu ( t^2/U )$ where a correction due to superexchange has been included.
Then the criteria for triplon condensation is given by
\begin{equation}
t^2/U = - \frac{ \lambda / 2}{\mathrm{min}( \phi_q^a \pm \phi_q^b ) + \nu}
\label{singleTriplonResult}
\end{equation}
\begin{equation}
\phi_q^a = \sum_\delta a_\delta \cos (\veck{q} \cdot \veck{\delta} )
\end{equation}
where ``$\mathrm{min}$'' refers to the most negative value of the argument and corresponds to the lowest part of the triplon energy band in Fig.~\ref{CondensationMechanismFigure}(a).
The extra $\nu$ term allows for the center of the triplon band to shift in energy with superexchange.
Since $t^2/U$ is positive by definition, $\mathrm{min}( \phi_q^a \pm \phi_q^b ) + \nu$ must be negative for condensation to be possible.

\section{Condensation from Spin-Orbital Superexchange}
\label{condensationSuperexchangeSupp}
We apply our formalism from Section~\ref{CalcOfCondHam} to diagonal hopping so that $t_{m_1 m_2} = 0$ if $m_1 \neq m_2$ to quantitatively obtain the key result of Section~\ref{ResForNorb321}.
This includes the three special cases ($N_\mathrm{orb} = 3,2,1$) from before, but generally applies to systems with corner sharing and face sharing octahedra.
The previous section showed expressing the Hamiltonian in the form of \eqref{generalMultipoleForm} had an immediate connection to the condensation Hamiltonian.
In this section, we explicitly calculate the coefficients $\alpha_{M}^{ll'L}$ and $\beta_{0}^{ss0}$ in \eqref{generalMultipoleForm} from hopping matrices to determine when the $^4 S$ pathway allows a F condensate.
For the scope of this section, we will make the additional simplification to throw away single site orbital anisotropy in the superexchange Hamiltonian (ie. terms like $L_{i,z}^2 + L_{j,z}^2$).
Using this condition and the restriction that the Hamiltonian must preserve time reversal symmetry, we are left to calculate $\alpha_{0}^{llL}$ and $\beta_{0}^{ss0}$ in \eqref{generalMultipoleForm} for arbitrary $l$, $L$, and $s$ for diagonal hopping matrices.
These conditions allow us to correctly guess the coefficients simply using conservation of angular momentum without resorting to more involved formalisms.

In second order perturbation theory, there are two applications of the $H_\mathrm{t}$ operator.
The first application contributes an angular momentum $k$ with amplitude $t_0^{k}$ while the second application contributes $k'$ with amplitude $(t^\dagger)_0^{k'}$.
Together this angular momentum is shared between $(L_i)_m^l$ and $(L_j)_{m'}^{l}$ and can be recast in terms of a total angular momentum using \eqref{orbitalAngularMomentumFromLiLj}.
The resulting orbital coefficient $\alpha_0^{llL}$ will be proportional to $\langle k 0, k ' 0 | L 0 \rangle$.

We denote the intermediate multiplet for the site  which gives up an electron during the first virtual hop ($^4 S$, $^2 D$, $^2 P$) as having orbital angular momentum $\xi_1$ and spin $\xi_2$ so that the multiplet is expressed as $^{2\xi_2 + 1} (\xi_1)$.
Then the product of the coefficients $\alpha_0^{llL} \beta_0^{ss0}$ is given below.
\begin{equation}
\alpha_0^{llL} \beta_0^{ss0} = 12 \, \tilde{\alpha}_0^{llL} \tilde{\beta}_0^{ss0} \, \langle p^2(1,1);p(\tfrac{1}{2},1) | p^3(\xi_2, \xi_1) \rangle^2
\end{equation}
Here $\langle p^2(S',L');p(S'',L'') | p^3(S''', L''') \rangle$ are the coefficients of fractional parentage\cite{Racah1943} for a $p$ shell and account for the total angular momenta being composed of identical particles while $\tilde{\alpha}_0^{llL}$ and $\tilde{\beta}_0^{ss0}$ are quantities which only depend on the recoupling of orbital and spin angular momentum previously described.
\begin{multline}
\tilde{\alpha}_0^{llL} = 
3
(-1)^{k' + \xi_1 + 1}
(2\xi_1 + 1)
(2l+1)^2
\langle k 0, k ' 0 | L 0 \rangle
\\
\times
(t^\dagger)_0^k (t)_0^{k '}
\left\lbrace \begin{array}{ccc} 1 & l & 1 \\ 1 & \xi_1 & 1 \end{array} \right\rbrace
\left\lbrace \begin{array}{ccc} 1 & k & 1 \\ l & L & l \\ 1 & k ' & 1 \end{array} \right\rbrace
\left\lbrace \begin{array}{ccc} 1 & l & 1 \\ 1 & 1 & 1 \end{array} \right\rbrace
\label{alphaExpression}
\end{multline}
The new symbol in braces is a Wigner-9j symbol.
The 9j symbol and the Clebsh-Gordan coefficient together contain the information of of how the angular momenta in the hopping matrix $t$ adds to give the orbital angular momenta in the resulting orbital part $\mathcal{O}_0^{llL}$ of the superexchange Hamiltonian.
The two 6j symbols along with the coefficients of fractional parentage contain the information of how angular moment is transferred between sites through the transfer of identical particles.
Lastly, our spin part is given similarly except that the restriction $S=\Sigma=0$ allows some simplification.
\begin{multline}
\tilde{\beta}_0^{ss0} = 
3
(-1)^{\xi_2 + 3/2}
(2\xi_2 + 1)
(2s+1)^{2}
\\
\times
\left\lbrace \begin{array}{ccc} \tfrac{1}{2} & s & \tfrac{1}{2} \\ 1 & \xi_2 & 1 \end{array} \right\rbrace
\frac{1}{\sqrt{2s+1}}
\left\lbrace \begin{array}{ccc} \tfrac{1}{2} & s & \tfrac{1}{2} \\ 1 & \tfrac{1}{2} & 1 \end{array} \right\rbrace
\label{betaExpression}
\end{multline}
With the coefficients $\alpha \beta$ determined, we can now use \eqref{generalMultipoleHopping} to determine when the ferromagnetic mechanism mediated by the $^4 S$ channel ($\xi_1 = 0$, $\xi_2 = \frac{3}{2}$) is active.

\end{document}